\documentclass[twocolumn,aps,tightenlines,floatfix,amsmath,showpacs]{revtex4}
\usepackage{graphicx,mathtools}
\usepackage{color}

\graphicspath{{pict/}}

\begin{document}

\title{Nuclear Level Density: Shell Model {\sl vs} Mean Field}

\author{Roman Sen'kov$^{1}$\footnote{email: rsenkov@lagcc.cuny.edu} and Vladimir Zelevinsky$^{2}$\footnote{email: zelevins@nscl.msu.edu}}

\affiliation{
$^{1}$Department of Natural Sciences, LaGuardia Community College, City University of New York, 31-10 Thomson Ave., Long Island City, NY 11101, USA\\
$^{2}$Department of Physics and Astronomy and National Superconducting Cyclotron Laboratory,
Michigan State University, East Lansing,
Michigan 48824-1321, USA\\
}


\begin{abstract}
The knowledge of the nuclear level density is necessary for understanding various reactions including those in the stellar environment. Usually the combinatorics of Fermi-gas plus pairing  is used for finding the level density. Recently a practical algorithm avoiding  diagonalization of  huge matrices was developed for calculating the density of many-body nuclear energy levels with certain quantum numbers for a full shell-model Hamiltonian. The underlying physics is that of quantum chaos and intrinsic thermalization in a closed system of interacting particles. We briefly explain this algorithm and, when possible, demonstrate the agreement of the results with those derived from exact diagonalization. The resulting level density is much smoother than that coming from the conventional mean-field combinatorics. We study the role of various components of residual interactions in the process of thermalization,  stressing the influence of incoherent collision-like processes. The shell-model results for the traditionally used parameters are also compared with standard phenomenological approaches.
\end{abstract}

\maketitle

\section{Introduction}

The knowledge of the level density is an important element in understanding the behavior of a quantum many-body system of interacting particles in various physical processes. In nuclear physics, this knowledge is necessary for the description of numerous reactions, including those of astrophysical or technological interest. The cross sections can be very sensitive to the level density that typically grows exponentially as a function of the excitation energy and the number of constituents. In turn, the theoretically predicted level density is sensitive to the statistics of particles, their specific interactions and available orbital space that, in realistic computation, usually has to be truncated. Apart from that, the level density in a finite self-bound system, such as the atomic nucleus, can be different for the available classes of eigenstates characterized by different quantum numbers of
exact constants of motion (in nuclei total spin $J$, parity $\Pi$ if we neglect weak interaction, and isospin $T$ if we neglect interactions violating charge-independence).

Below we consider a problem of practical microscopic calculation of the level density for a nucleus described by a Hamiltonian of the shell-model type. In this framework it does not matter if the Hamiltonian is derived from a more fundamental approach or fit phenomenologically with the use of experimental data. We assume that the Hamiltonian describes the  low-lying energy spectra, transition rates and other observables known from the experiment reasonably well. Then the task is to predict the level density of the system
at higher excitation energy, in the region beyond direct measurements resolving individual quantum states. Practically, the relevant factual milestones, apart from the low-lying spectroscopy,  are the regions of isolated neutron resonances near neutron separation energy and the results of the Oslo method and related experimental approaches  \cite{schiller00,voinov07,larsen11,egidy12}. Of course, any practical shell-model Hamiltonian loses its validity outside of the truncated orbital space where this Hamiltonian was expected to work. At some excitation energy, the states of the system come from the particle orbitals not included in the model. However, with available computational means, the space of validity of the model Hamiltonian can be broad enough, in particular including the excited states involved, for example, in astrophysical reactions at a typical stellar temperature. We can also hope to use the microscopic results for the nuclei far from stability, where the level density is usually predicted by pure phenomenology \cite{kawano06}.

Another physical limitation arises from the obvious fact that the states involved in the reactions belong to the continuum, while the standard shell-model calculations work in the discrete spectrum. Instead of discrete states here one has to deal with resonances seen in various reactions. Then the whole definition of the level density becomes questionable and, strictly speaking, one has to move to the complex plane of resonances. However, the traditional approach is still meaningful if the typical widths of the involved states are small compared to the spacings between the states with the same quantum numbers. In what follows we limit ourselves by this situation.

Neglecting the continuum effects, the trivial solution for the level density generated by a certain Hamiltonian would be a full diagonalization of the Hamiltonian matrix in an appropriate orbital basis. However, this is practically only possible in sharply truncated orbital spaces, which might be only sufficient for relatively light nuclei, like in the $sd$-shell \cite{big}. In many realistic cases of current interest, the dimensions of corresponding matrices, even in subspaces with given quantum numbers, are prohibitively large. Moreover, such a diagonalization is anyway superfluous because we do not need full information on every excited state in spectral regions of high level density. The level density is, by construction, a statistical notion.

In this situation, we are looking for the statistical solution of the problem.
The physical justification of such an approach lies in the fact that, at small level spacings, the stationary nuclear states are extremely complicated superpositions of simple determinantal states with integer occupation numbers of definite orbitals. Gradually switching on inter-particle interactions and going in this process through multiple avoided crossings of various configurations we come to {\sl chaotic} states \cite{big,perc} with observable properties smoothly changing along the spectrum. Therefore our problem reduces to finding a realistic way to describe this smooth evolution.

This purpose can be reached using the methods of {\sl statistical nuclear spectroscopy} \cite{FR,wong86,kota10}. Already in the framework of a single partition (a certain configuration of independent particles occupying the mean field levels), the level density after including the particle interaction rapidly goes to the Gaussian limit with the increasing particle number \cite{brody81}. This is some kind of manifestation of the central limit theorem. The many-level, and therefore many-partition, generalization should give a reliable image of the total level density for an accepted orbital scheme. This has to be done for each class of global quantum numbers. This direction of theoretical search has a long history. We would like especially mention the works in the direction of statistical spectroscopy applied to shell-model Hamiltonians, see for example \cite{grimes79,johnson01}. After several preliminary publications, our successful algorithm was constructed \cite{PLB11} and opened for public use \cite{CPC13}. The results of implementing this algorithm for the level density in sectors with given values of global constants of motion are practically identical to those from the full diagonalization when the latter is possible. For well-tested  shell-model Hamiltonians, the results are in good agreement with available experimental data.

For many years, starting with the classical work by Bethe \cite{bethe36}, the nuclear level density was estimated using the combinatorics based on the ideas of a Fermi gas. The influential review of earlier approaches of this type was given by Ericson \cite{ericson60}, the later derivations can be found in \cite{gilbert65,baba70,holmes76,engelbrecht91}, see also \cite{BM}. The recent achievements in this direction \cite{gorieli08, uhrenholt09, hilaire12} include the pairing correlations considered as a part of the self-consistent mean field in the framework of the BCS theory or Hartree-Fock-Bogoliubov variational ansatz. The shell-model Monte-Carlo methods  \cite{alhassid00,alhassid07,alhassid12}, being very demanding computationally, work relatively well at least with some parts of the full shell-model interaction but require the projection to the correct values of spin and parity.

The chaotization of the dynamics mentioned above leads to the possibility of describing the physics of excited states at high level density in terms of statistical thermodynamics including temperature, entropy etc. This was understood in application to nuclear reactions from the early times of nuclear physics \cite{bethe36,frenkel36,landau37}. The detailed analysis of atomic \cite{flambaum94,FI97} and nuclear \cite{big,ann} chaotic states supported an old idea \cite{LL} of thermalization in a closed system driven by the interactions between the constituents, with no heat bath: the average over a generic chaotic wave function in a chaotic region is equivalent to the average over a standard equilibrium thermal ensemble \cite{srednicki94}. Currently this idea, sometimes called the ``eigenfunction thermalization hypothesis'', is extensively discussed in the many-body physics community \cite{polkovnikov11}. One of the purposes of the current publication is in comparison of the exact shell-model nuclear level density with phenomenological ideas based on the Fermi-gas picture at certain temperature. We look at these ideas and based on them equations from the viewpoint of our numerical results. Our attention will be mostly concentrated on the usually cited empirical parameters of the level density and their energy and spin dependence. Another point of interest is in the role of various components of the shell-model interactions in the formation of the level density. One important result is that the consideration of the mean field, even with addition of the BCS-type pairing, is not sufficient. Incoherent components of the interaction in a finite many-body system, as a rule neglected in the mean-field combinatorics, play a significant role smoothing the energy behavior of the level density.

In what follows we briefly explain the method and give the examples of practical calculations. The results will be compared with what would follow from traditional phenomenological models.

\section{Moments method}

We consider a finite system of interacting fermions described by the standard Hamiltonian
\begin{equation}
 H=\sum_{1}\epsilon_{1}a^{\dagger}_{1}a_{1}+\,\frac{1}{4}\,\sum_{1234}V_{12;34}a^{\dagger}_{1}a^{\dagger}_{2}a_{3}a_{4}         \label{1}
\end{equation}
that contains the mean-field part with effective single-particle energies $\epsilon_{1}$ and the antisymmetrized two-body interaction. The generalized numerical subscripts combine all quantum numbers of single-particle orbitals. In this form the method can be applied to nuclei, atomic or molecular electrons, atoms in traps etc., any system where the residual interaction is sufficiently strong to produce complicated eigenstates. The three-body forces can be included in the same way although the computations become more cumbersome. Using a phenomenological shell-model Hamiltonian we assume that many-body forces, at least partly, are included in fitted matrix elements. The quality of the Hamiltonian is checked by the explicit applications to individual low-lying states and comparison with available experimental information.

In a finite self-bound system, such as the atomic nucleus, the total angular momentum  (nuclear spin) is exactly conserved supplying good global quantum numbers $J$ and $M$. Therefore it is convenient from the very beginning to use a spherically symmetric basis of single-particle orbitals $|jm)$ which define, along with the orbital momentum $\ell$, main quantum number $\nu$, and isospin $\tau$, the quantum numbers combined in Eq. (\ref{1}) into a unified numerical subscript. If the orbital space is sufficiently broad, this spherical shell model can describe intrinsic deformation without violating rotational symmetry \cite{BZ}. In practice, the operators in the Hamiltonian can be combined into pairs, such as $(a_{3}a_{4})_{L\Lambda}$ with the angular momentum quantum numbers of the pair, $L\Lambda$, fixed through the Clebsch-Gordan coefficients. In the same way, for the isospin-invariant forces, the isospin of the pair can be also fixed, and then the interaction $V$ in a restricted orbital space is defined through a finite number of pairwise matrix elements. If more convenient for computations, one can as well  use the $M$-scheme without vector coupling.

The practical algorithm of calculations in the $M$-scheme follows from individual configurations, $p$, ({\sl partitions}) which are possible distributions $(n_{1},n_{2},...)$ of available particles, $\sum_{1}n_{1}=N$, over single-particle orbitals (here the index {\sl 1} does not include the projection $m$). The many-body states $|\alpha\rangle$ possible for each partition form a subspace where $\alpha$ combines total quantum numbers $N,M,T_{3}$ and parity. It is convenient \cite{PLB11} to use the proton-neutron formalism.

Let $D_{\alpha p}$ be the dimension of the class of states with global quantum numbers $\alpha$ built on the partition $p$. As shown in statistical spectroscopy \cite{brody81,wong86} and confirmed in many examples by the exact shell-model diagonalization, the density of states for a given partition is close to the Gaussian. Of course, this is the main assumption based on a rich experience with the features of quantum chaos in mesoscopic systems. The characteristics of the Gaussian are defined by the moments (traces) of the actual Hamiltonian. The centroid is just the mean energy value for a given partition,
\begin{equation}
E_{\alpha p}={\langle H\rangle_{\alpha p}}=\,\frac{1}{D_{\alpha p}}
\,{\rm Tr}^{(\alpha p)}H.                                                                          \label{2}
\end{equation}
The dispersion of the Gaussian, $\sigma_{\alpha p}$, is the second moment,
\begin{equation}
\sigma_{\alpha p}^{2}={\langle H^{2}\rangle_{\alpha p}}-E_{\alpha p}^{2}\equiv\,
\frac{1}{D_{\alpha p}}\,{\rm Tr}^{(\alpha p)}H^{2} -E_{\alpha p}^{2}.                     \label{3}
\end{equation}
It is important to stress that the calculation of these traces does not require the diagonalization of large-scale matrices. The first moment (\ref{2}) is the diagonal matrix element of the Hamiltonian  averaged over the partition, while the second moment (\ref{3}) is the sum of squared off-diagonal  elements along one line of the Hamiltonian matrix, again averaged over the lines corresponding to the partition. It is known that the dispersion for each basis state very weakly fluctuates within a partition \cite{big,pillet12} even prior to the  next averaging. The second moment includes all interactions coupling the partitions. If the traces were calculated in the $M$-scheme, we obtain the {\sl density of states} counting all $M$-degenerate states within the multiplets. To obtain the {\sl level density} for given spin $J$, we have in a standard way to find the difference of traces for $M=J$ and $M=J+1$.

\begin{figure*}
\begin{tabular}{cc}
\includegraphics[width=0.4\textwidth]{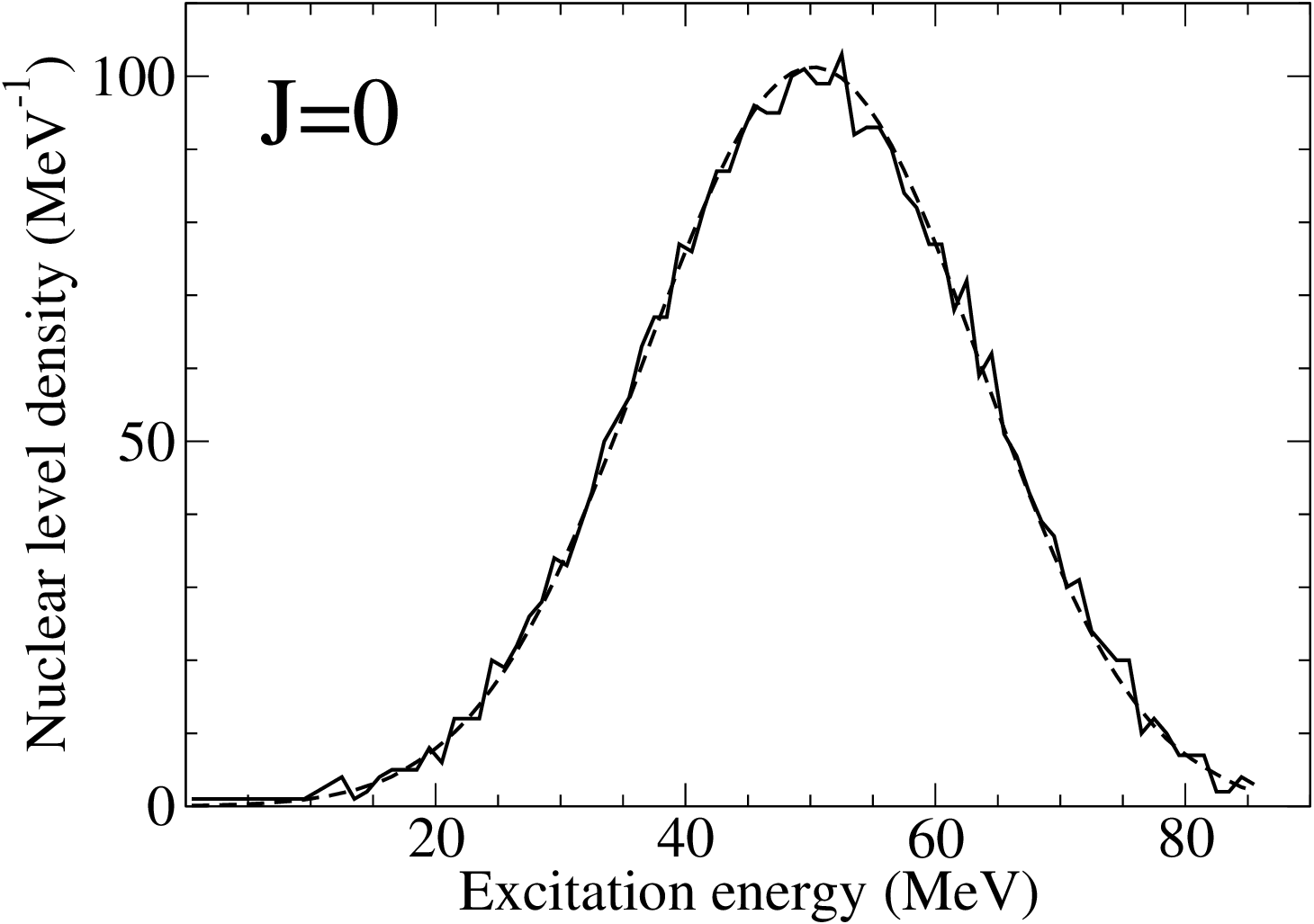} \hspace*{0.5cm} &
\includegraphics[width=0.4\textwidth]{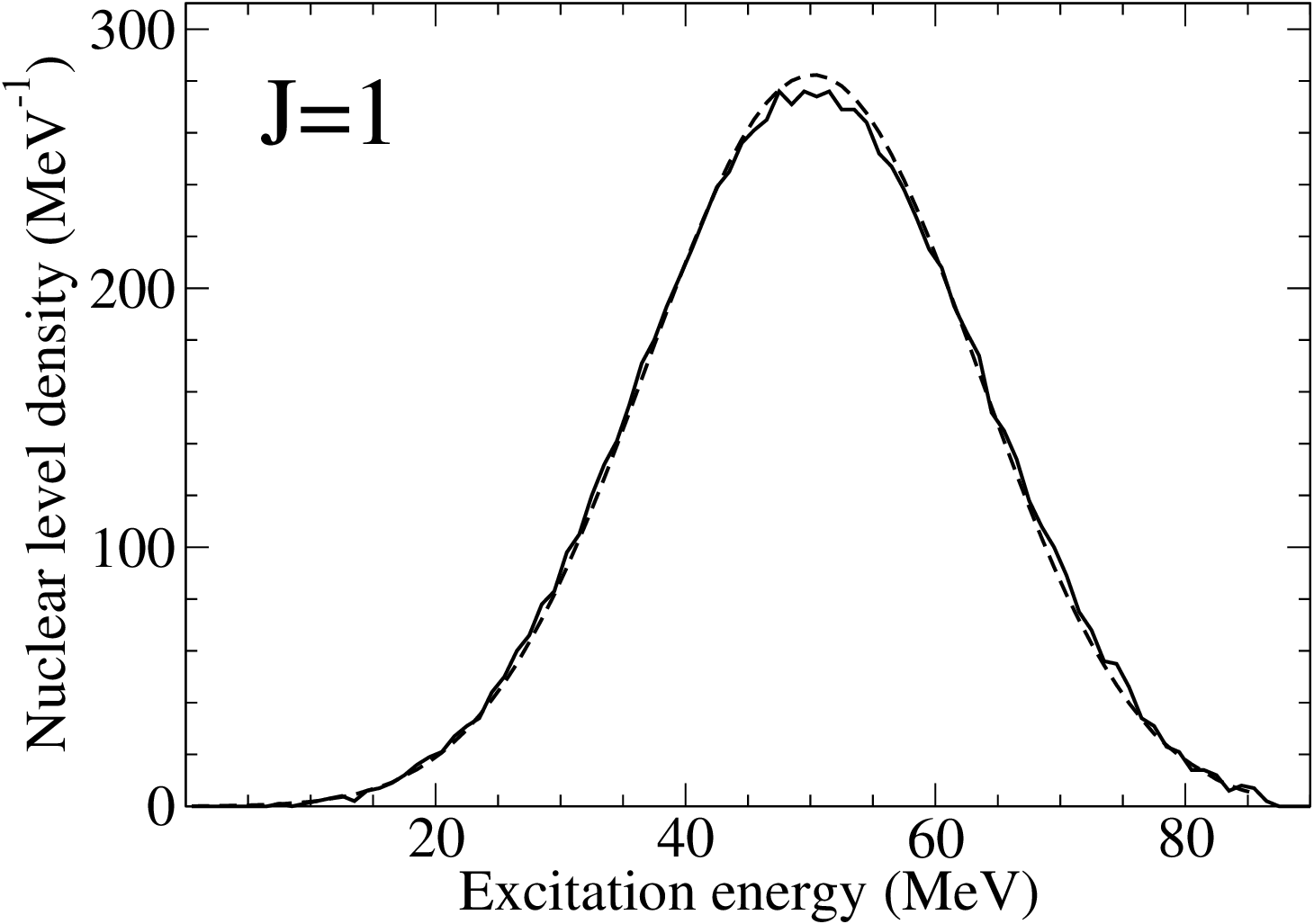}
\vspace*{0.5cm} \\
\includegraphics[width=0.4\textwidth]{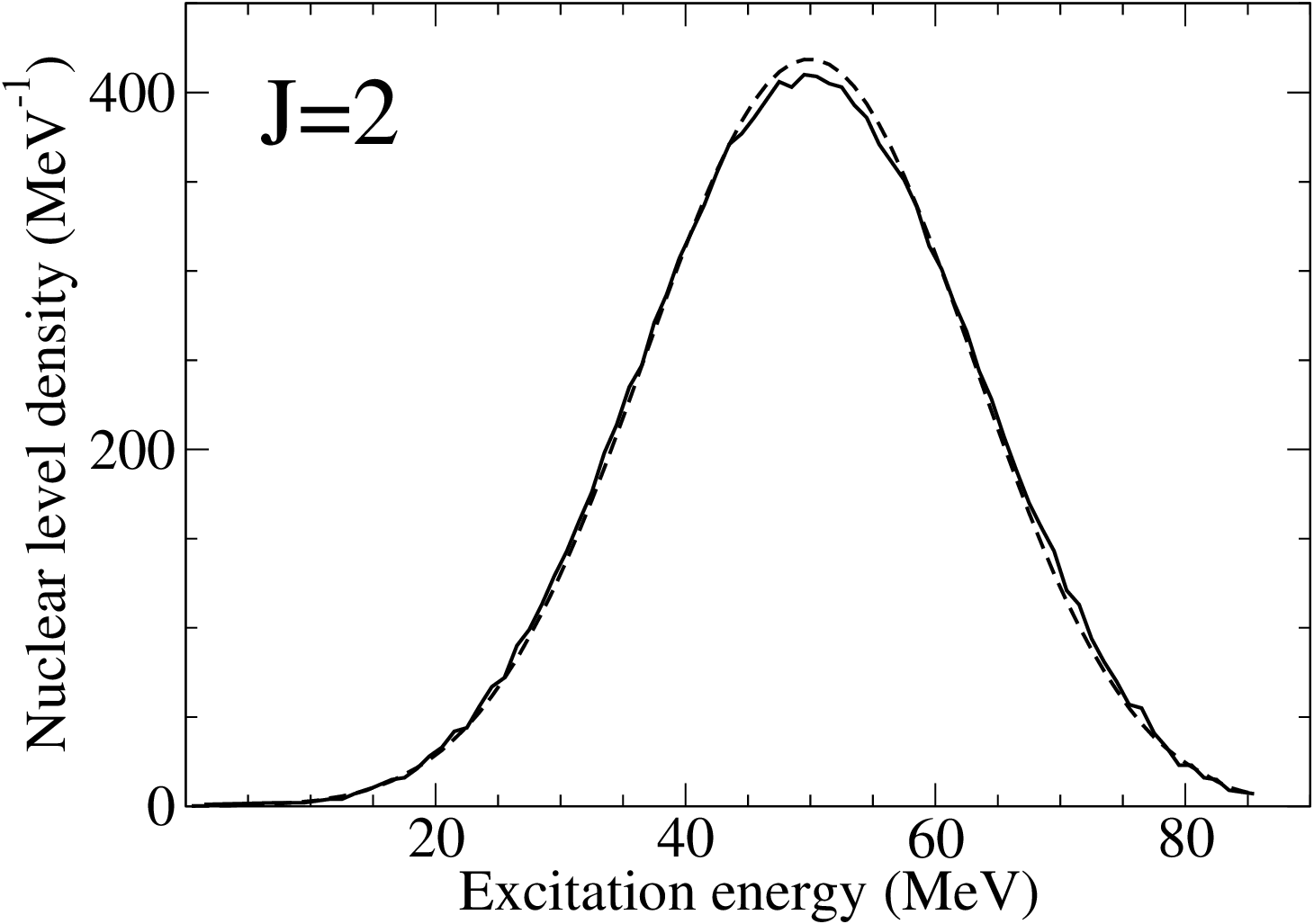} \hspace*{0.5cm} &
\includegraphics[width=0.4\textwidth]{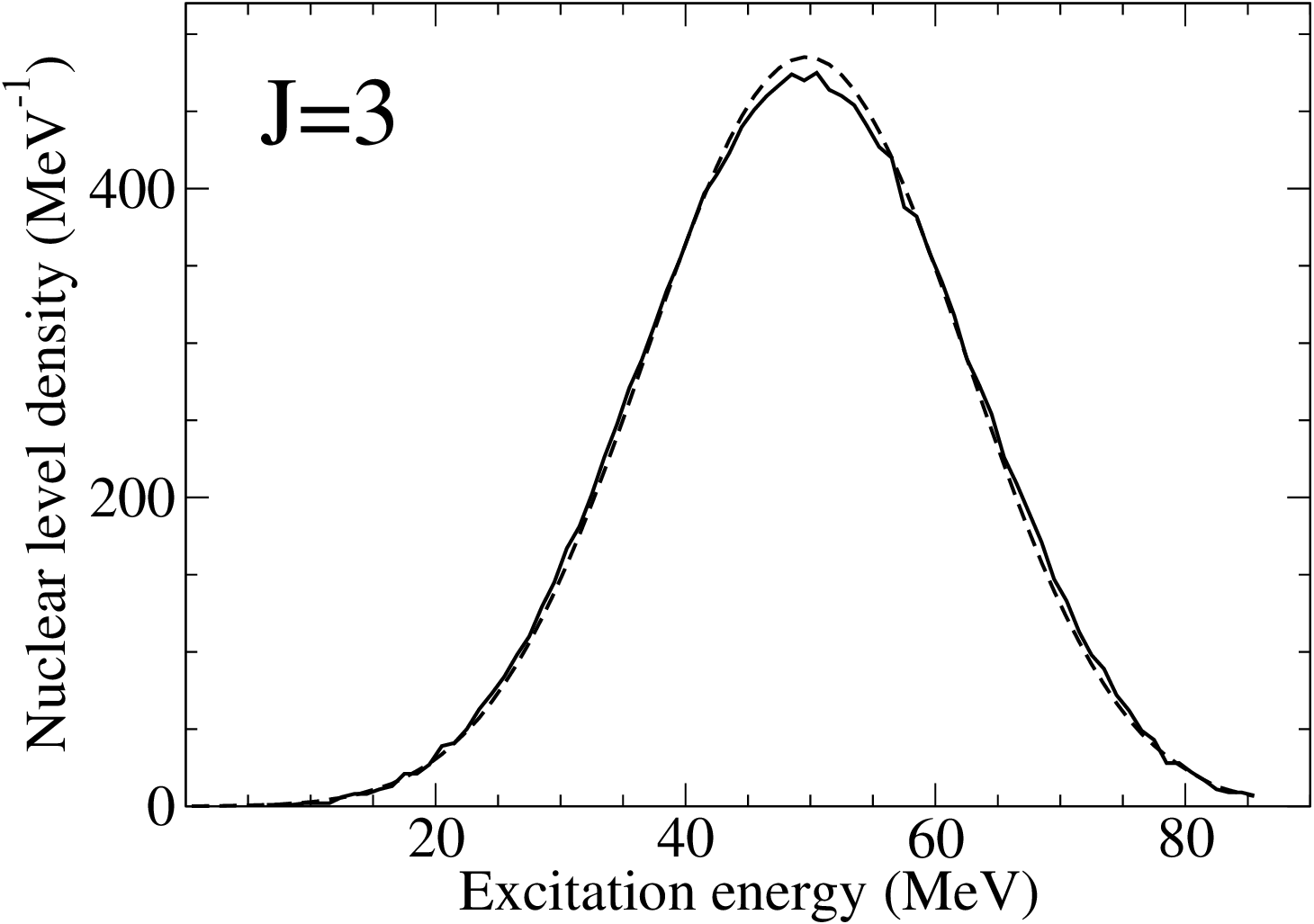}
\end{tabular}
\caption{ Nuclear level densities for $^{28}$Si, $\Pi=+1$, various spins;
shell model  for the $sd$-shell and USDB two-body interaction (solid curves) {\sl vs.} moments method (dashed curves); finite-range  parameter $\eta=2.8$.\\}
\label{fig1}
\end{figure*}

The total level density is given by summing the contributions of partitions using the constructed Gaussians $G_{\alpha p}(E)$ with their centroids (\ref{2}) and widths (\ref{3}),
\begin{equation}
\rho(E;\alpha)=\sum_{p}D_{\alpha p}G_{\alpha p}(E).                          \label{4}
\end{equation}
As was understood earlier \cite{kaiser03}, it is better to use {\sl finite range Gaussians},
\begin{equation}
G_{\alpha p}(E)=G(E-E_{\alpha p}+E_{{\rm g.s.}};\sigma_{\alpha p}),             \label{5}
\end{equation}
where
\begin{equation}
G(x;\sigma)={\cal C} \left\{
\begin{array}{ll}
e^{-x^{2}/2\sigma^{2}}, & |x|\leq \eta\sigma \\
0 , &  |x|>\eta\sigma
\end{array}\right. .                                                            \label{6}
\end{equation}
Here $E_{{\rm g.s.}}$ is the ground state energy to be defined separately, ${\cal C}$ is the normalizing factor, $\int dx\, G(x,\sigma)=1$, and the cutting finite-range parameter $\eta$ has to be found empirically \cite{SH10}; its actual value $\eta\approx 2.8$ agrees with the analysis of the shape of typical nuclear strength functions \cite{frazier}.

Following the recipe of Ref. \cite{HZ07} we make an important addition to the algorithm. In many realistic applications, the orbital space contains several shells labelled by the harmonic oscillator quanta ${\cal N}$. In this case, the standard formulation of the shell model includes cross-shell transitions with unphysical excitations of the center-of-mass. These spurious states are to be excluded from the level density. In some versions of the shell model, these states are artificially shifted to high energies. Here, the subtraction of ghost states is accomplished by renormalizing the contaminated level density $\rho(E,J;{\cal N})$ through the recurrence relations. For example, while  the ${\cal N}=0$ case, which will be called the $\rho^{\circ}(E,J;0)$ approximation, is free of admixtures, the pure level density $\rho^{\circ}$ at the next step (no admixtures of the single center-of-mass excitation) is found as
\begin{equation}
\rho^{\circ}(E,J;1)=\rho(E,J;1)-\sum_{J'\le |J-1|}^{J+1}\rho(E,J';0).                        \label{7}
\end{equation}
Here the sum goes over the intermediate angular momenta $J'$ from $|J-1|$ to $J+1$ since the center-of-mass operator is equivalent to a vector. If higher admixtures ${\cal N}>1$ are present, the recurrence relation has to include a corresponding number of steps back. This makes the calculation of the trace of $H^{2}$ with various intermediate states slightly more involved \cite{jacquemin81,SH10}.

\section{Examples of level density}

\subsection{Comparison with the exact solution of the shell model}

The first natural check of the approach is in comparison of the resulting level density with the picture arising from the full shell-model diagonalization in the cases where such a diagonalization is technically plausible. The $sd$-shell model for a long time is known as the best example of exact diagonalization. The model is completely fixed by the effective single-particle energies $d_{5/2}, s_{1/2}$, and $d_{3/2}$, and 63 phenomenologically fitted matrix elements of two-body interaction. The model works extremely well for practically all observables of $sd$-nuclei and not only for the lowest states. For example, both, the experiment \cite{brenn95}, and the $sd$-shell model
\cite{HZ07}, indicate the existence of  ten stationary states with $J^{\Pi}=0^{+}$ up to excitation energy of 15 MeV in $^{28}$Si, therefore providing the same average level density at least at not very high energy; the mean level spacing between those $0^{+}$ levels is 0.95 MeV in the experiment
and 1.02 MeV in the shell-model calculation.

\begin{figure*}
\begin{tabular}{cc}
\includegraphics[width=0.45\textwidth]{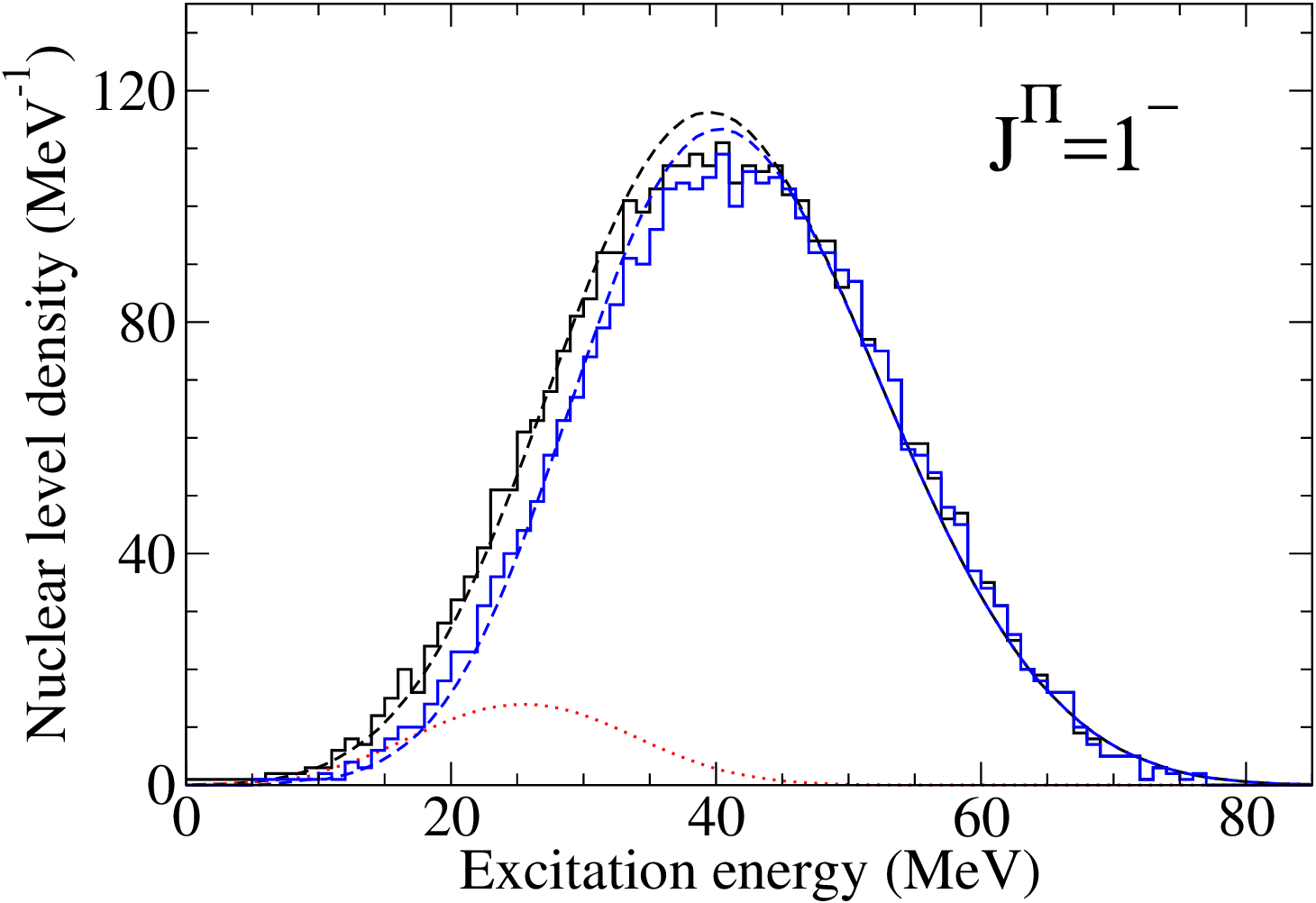} \hspace*{0.5cm} &
\includegraphics[width=0.45\textwidth]{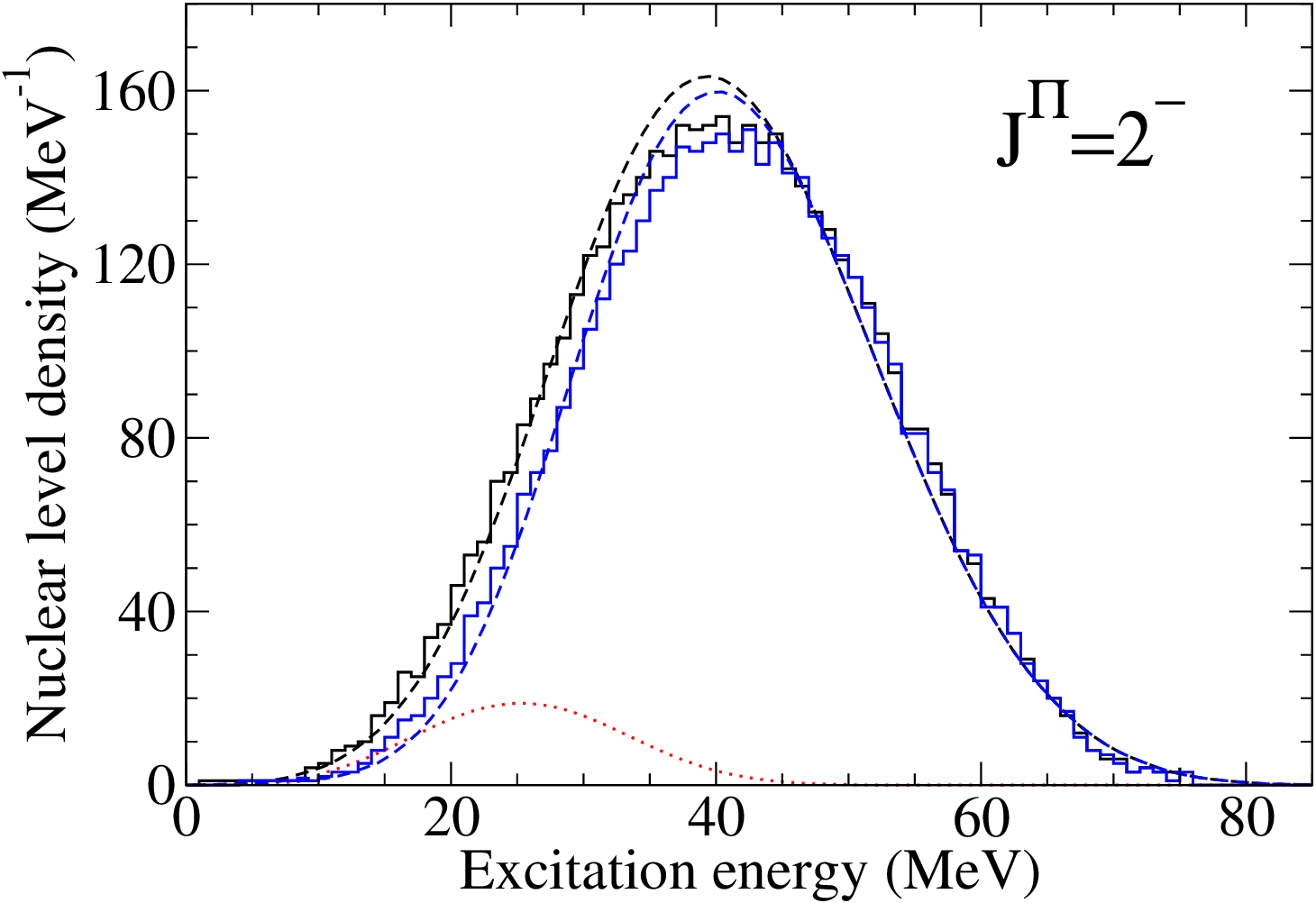}
\end{tabular}
\caption{(Color online.)  Comparison of the nuclear level densities of $^{20}$Ne in $1\hbar\omega$-approximation, $\Pi=-1$. The densities are calculated within the shell model approach (stair-case curves) and with the moments method (solid curves). The black curves on the left and right graphs (both are higher at low energies) correspond to the densities with spurious states included; the blue curves (that are lower at low energies) correspond to the densities without spurious states; the red (dotted) curves present the spurious nuclear level densities.}
\vspace*{0.5cm}
\label{fig2}
\end{figure*}

\begin{figure*}
\begin{tabular}{cc}
\includegraphics[width=0.45\textwidth]{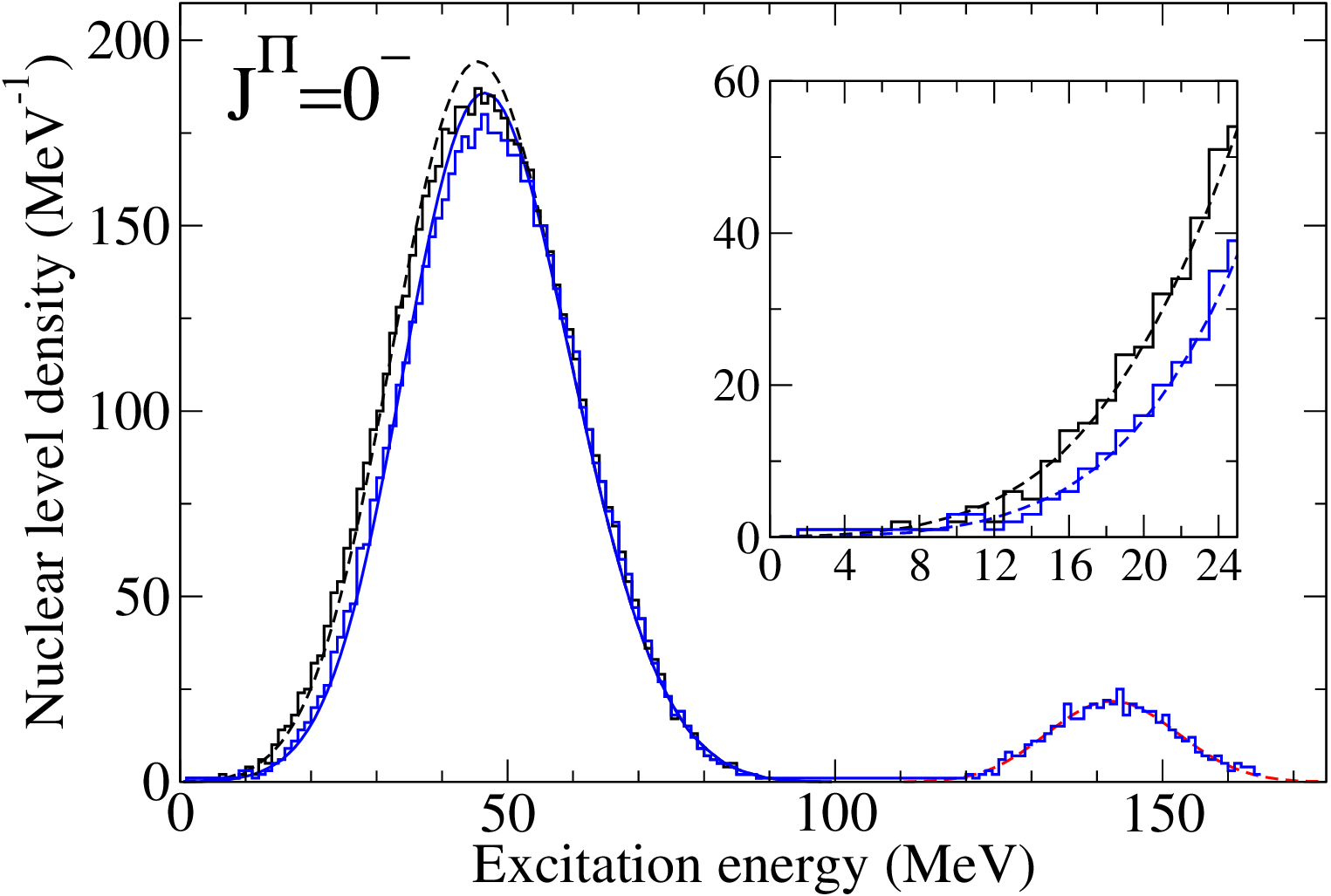} \hspace*{0.5cm} &
\includegraphics[width=0.45\textwidth]{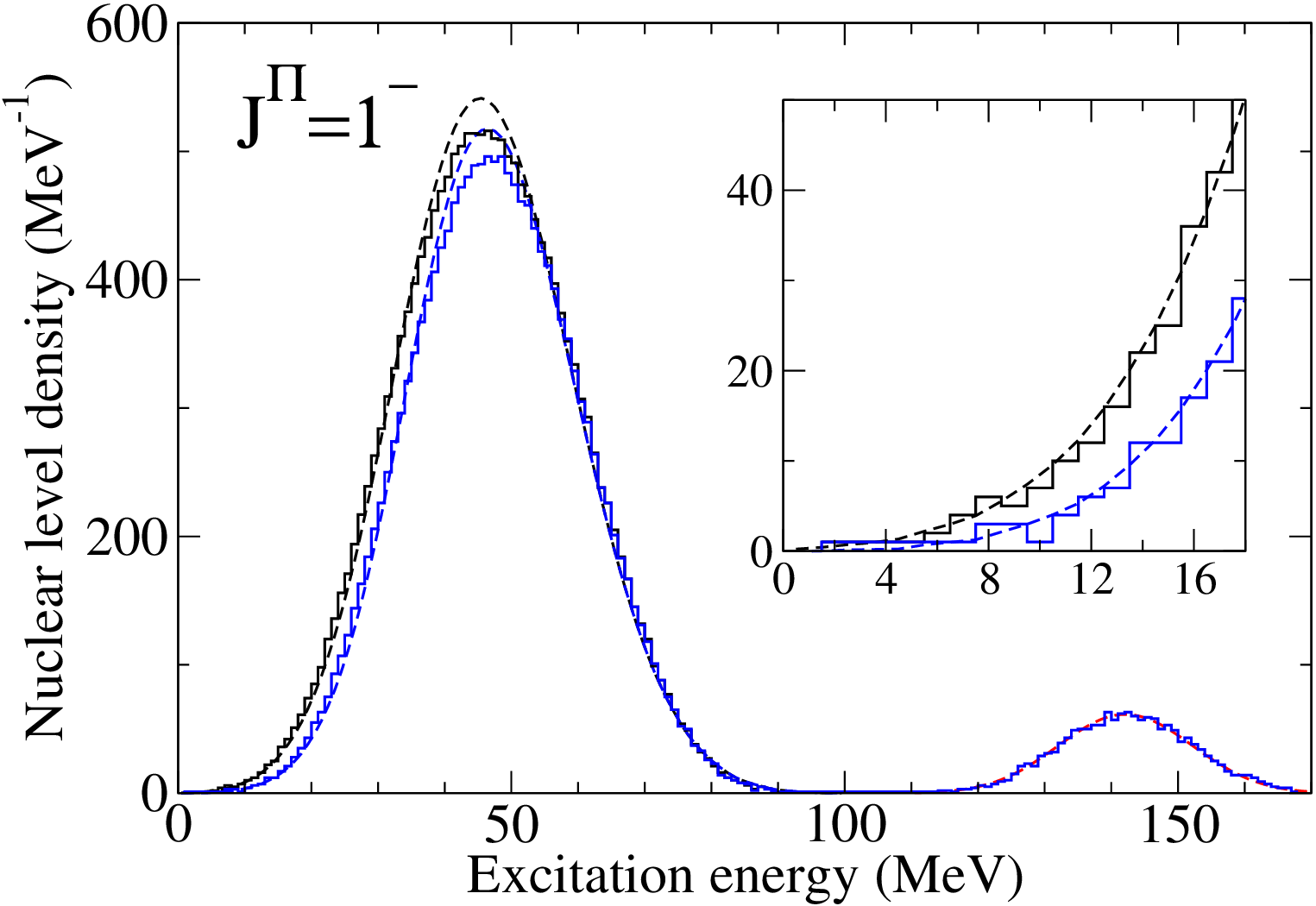}
\end{tabular}
\caption{(Color online.) Comparison of nuclear level densities calculated with the shell model for $^{22}$Mg in  $1\hbar\omega$-approximation, $\Pi=-1$, and the moments method. The coloring scheme is the same as in Fig. \ref{fig2}. The small peaks around 140 MeV show the spurious densities. The inserts present the low-energy part.\\}
\label{fig3}
\end{figure*}

Fig. \ref{fig1} illustrates the results of the moments method for calculating the level density in $^{28}$Si, a typical object of the $sd$-shell model applications that has served long ago as a testing ground for quantum chaos \cite{big}. We see that the level density for different classes of states,
here $0^{+}, 1^{+}, 2^{+},$ and $3^{+}$,  is always a smooth curve of the Gaussian type. Of course, as the calculations have been done in the restricted orbital space, the real physical result that can be juxtaposed to the experimental data and used for the reaction calculations always corresponds only to the left hand side of the full graph and to the  excitation energy below the centroid maximum.

Various specific quantum numbers (total spin and parity) produce the level density of the same qualitative behavior, with the integral corresponding to exact multiplicities of  states with a given set of quantum numbers in a fixed orbital space. All examples look the same. The agreement with the exact shell-model diagonalization is almost perfect, with slightly more visible fluctuations for the class $J^{\Pi}=0^{+}$ that has a smaller total dimension. In all cases we see a small deviation near the centroid which supposedly can be eliminated by taking into account the fourth moment of the Hamiltonian (but it makes no sense to go for such complicated and time-consuming calculations to improve the results in the region outside the physically relevant area).

The smoothness and Gaussian behavior of results in all cases confirm the possible thermodynamic interpretation in terms of  entropy $S(E)$ (mean logarithm of the level density) and temperature, $dS/dE=1/T(E)$. Formally, the centroid of the level density for the finite orbital space corresponds to infinite temperature and the right half of the curve to negative temperatures.
The full shell-model analysis of the wave functions \cite{big} has found that the same effective temperature can be extracted by the single-particle thermometer using the occupation numbers of available spherical orbitals {\sl for individual stationary many-body states}. The interaction of the quasiparticles in the self-consistent mean field acts as the heat bath, and the chaotic mixing of the eigenstates leads to thermalization even in such a small Fermi system.

\begin{figure*}
\begin{tabular}{cc}
\includegraphics[width=0.45\textwidth]{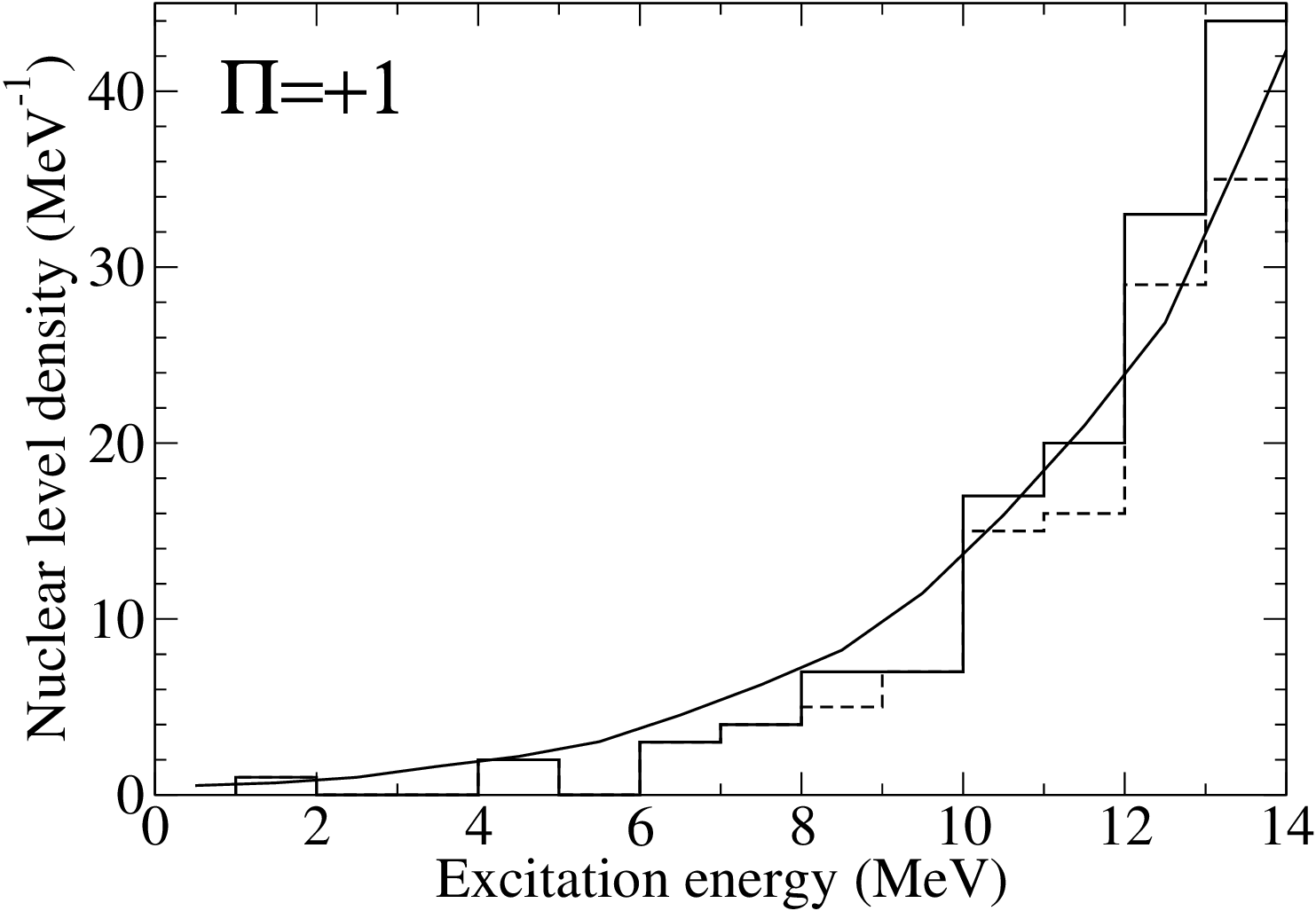} \hspace*{0.5cm} &
\includegraphics[width=0.45\textwidth]{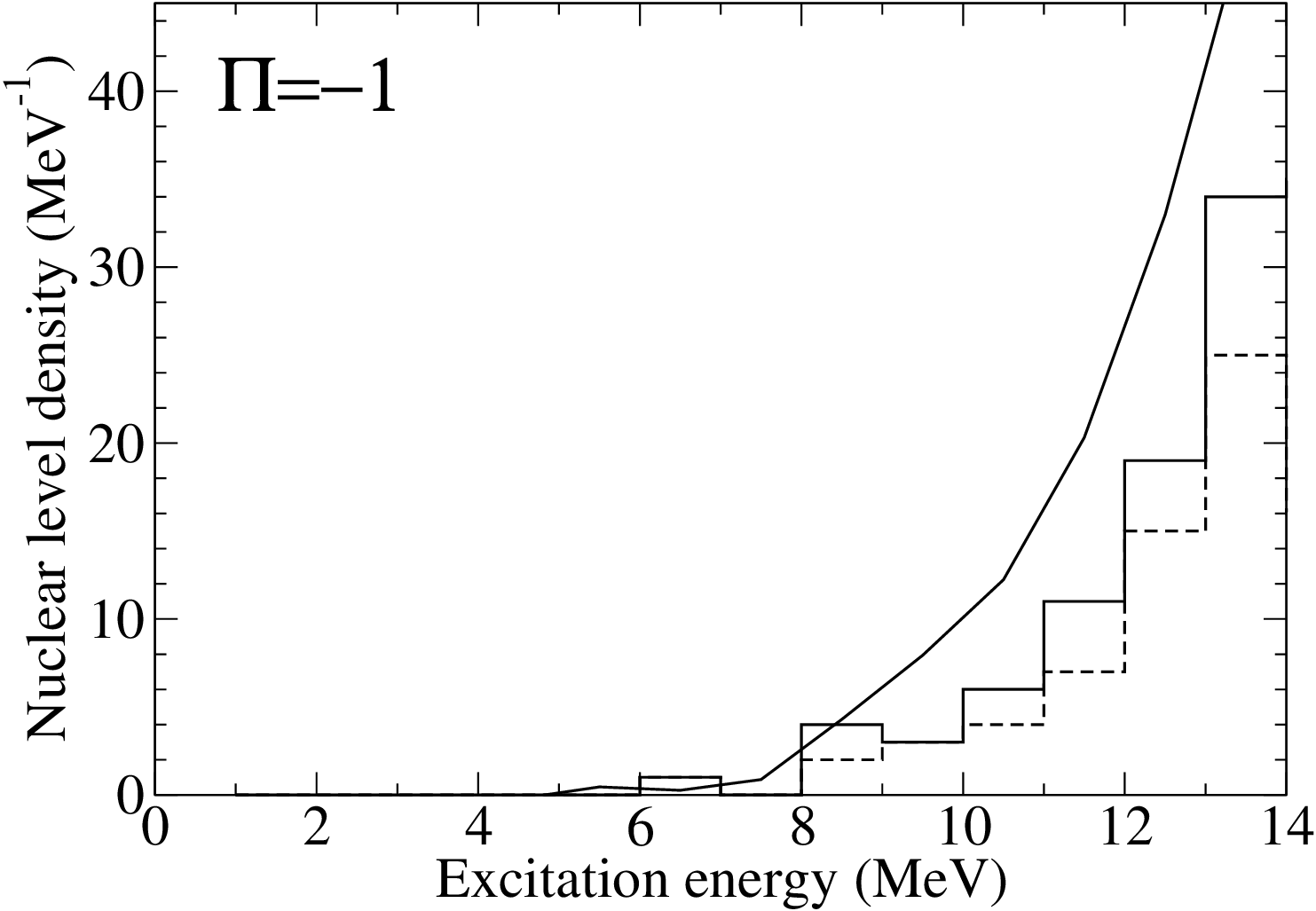}
\end{tabular}
\caption{Comparison of the experimental nuclear level densities (stair lines) and the densities calculated with the moments method (straight lines) for $^{28}$Si, all $J$, positive (the left graph) and negative (the right graph) parity. \\
\vspace*{0.3cm}}
\label{fig4}
\end{figure*}

\begin{figure*}
\begin{tabular}{cc}
\includegraphics[width=0.45\textwidth]{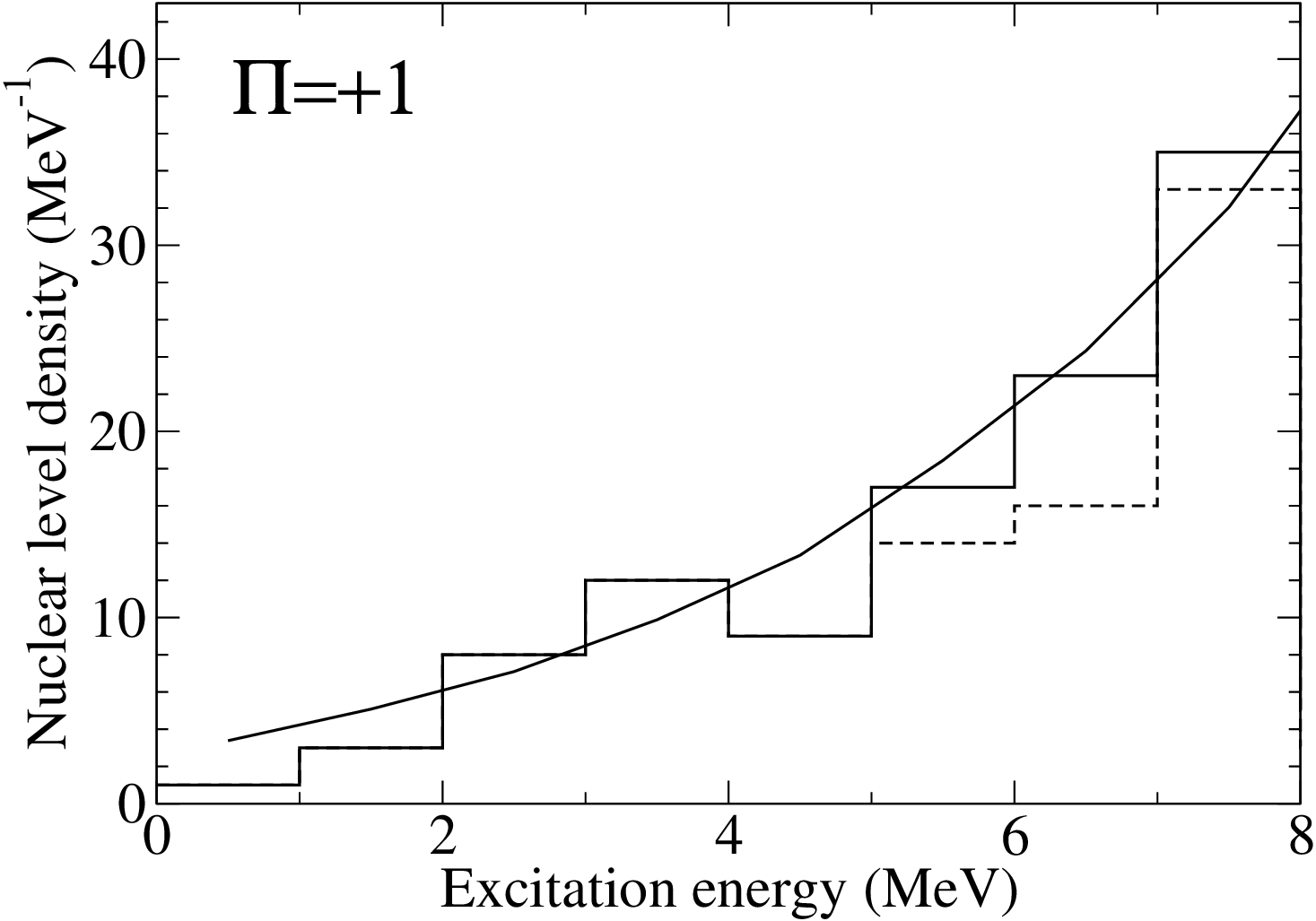} \hspace*{0.5cm} &
\includegraphics[width=0.45\textwidth]{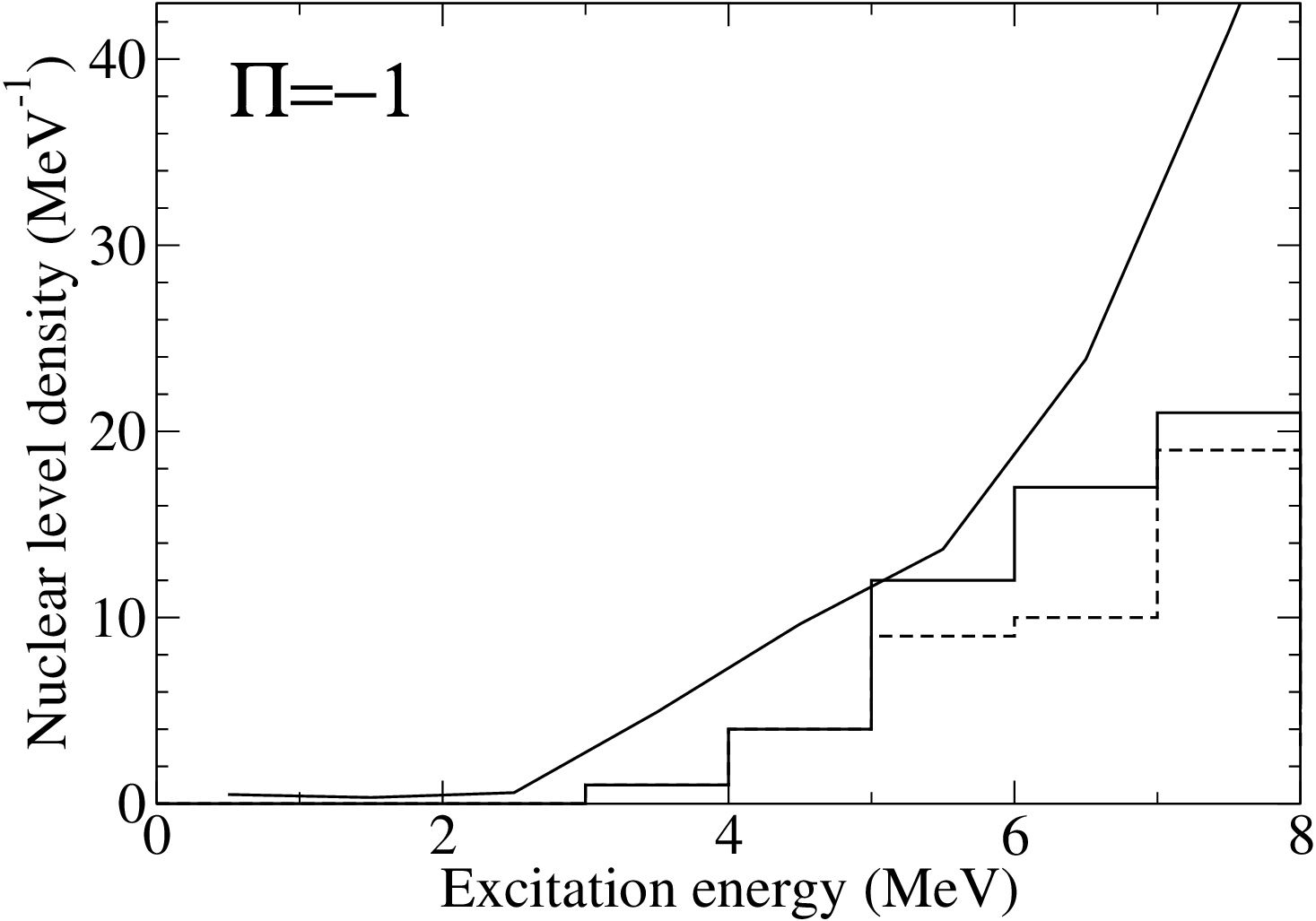}
\end{tabular}
\caption{Comparison of the experimental nuclear level densities (stair lines) and the densities calculated with the moments method (straight lines) for $^{26}$Al, all $J$, positive (the left graph) and negative (the right graph) parity. \\}
\label{fig5}
\end{figure*}

\subsection{Elimination of spurious states}

As mentioned above, the full shell-model diagonalization in the cases with the presence of transitions between the orbitals of opposite parity (excitations across the oscillator shells) brings in ghost states related to the center-of-mass motion rather than to intrinsic excitations. We explained above the recurrent techniques used for eliminating these spurious states and obtaining the pure level density. The recipe frequently used in the shell model is the brute-force shift of the undesired states to high energy by adding to the Hamiltonian under diagonalization a Lawson term \cite{lawson} that in the harmonic oscillator field of frequency $\omega$ looks as ($\beta>0$)
\begin{equation}
H'=\beta\left[H_{{\rm c.m.}}-\,\frac{3}{2}\,\hbar\omega\right]\,\frac{A}{\hbar\omega}.\label{8}
\end{equation}
As was shown long ago \cite{big}, this recipe indeed generates a new branch of eigenstates shifted to high energy (by about $\sim \beta {\cal N} A$) but having essentially the same complexity (measured by the information entropy) as their predecessors without spurious admixtures.

The separation of unphysical spurious states according to the recurrence relations (\ref{7}) works well. The standard shell-model shift of these states to higher energy produces exactly the same group of states as our procedure. Fig. \ref{fig2} shows the level density calculated for $^{20}$Ne in the broad space of $s+p+sd+pf$ orbitals in the $1\hbar\omega$ approximation (only one-step excitation of  negative parity states that include non-physical admixtures). The full (``blind") calculation compared to that treated with recurrence relations (\ref{7}) contains the excess shown by the small (red dotted) Gaussian-type curves at relatively low energy. We see that this curve of difference exactly coincides with the result of the shell-model shift according to Eq. (\ref{8}). This means that the method of recurrence relations correctly removes the unphysical excitations. The same conclusion can be made for $^{22}$Mg, Fig. \ref{fig3}, where the insets zoom out the region
of low energy.

\subsection{Comparison to experimental data}

\begin{figure*}
\begin{tabular}{cc}
\includegraphics[width=0.4\textwidth]{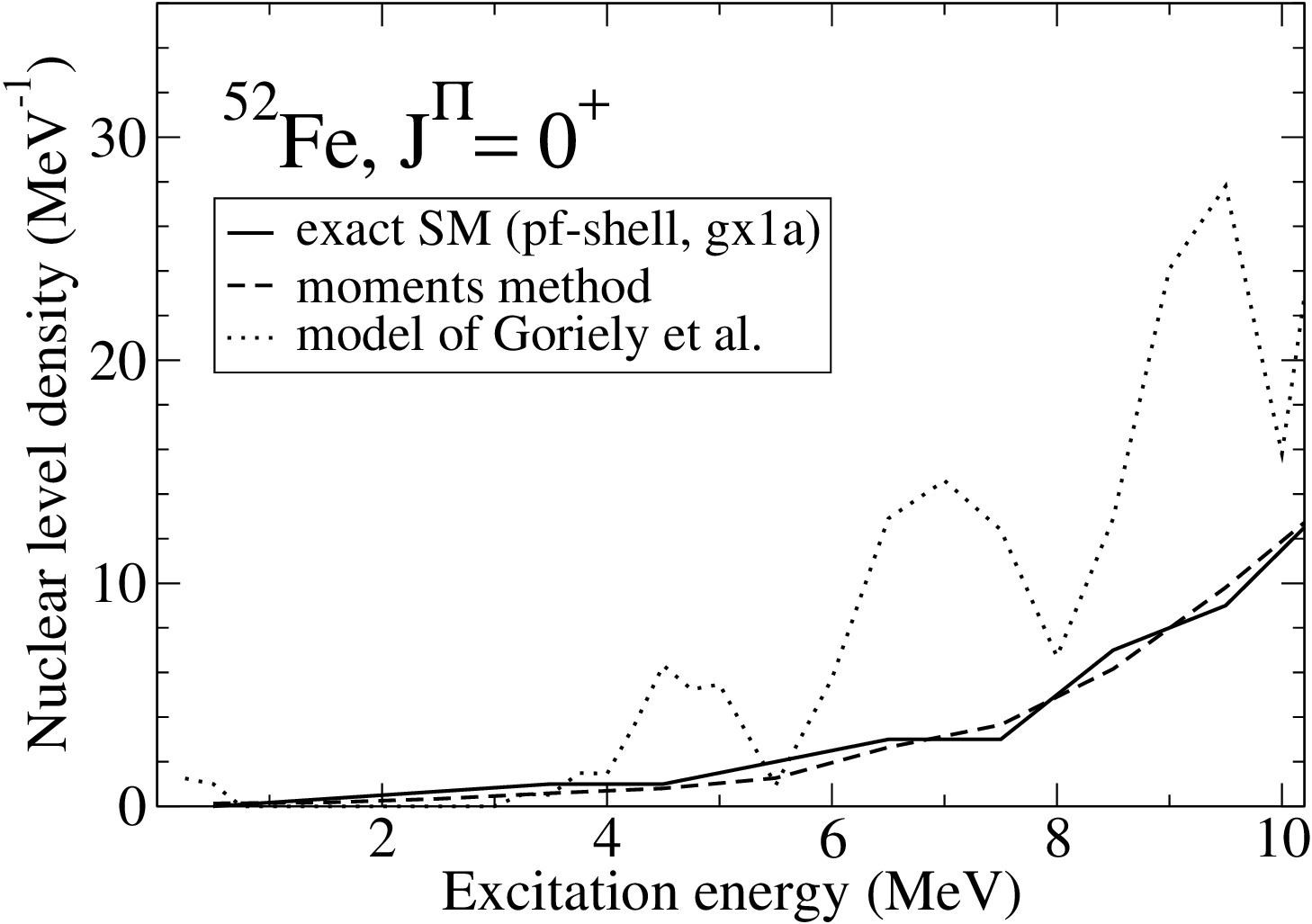} \hspace*{0.5cm} &
\includegraphics[width=0.4\textwidth]{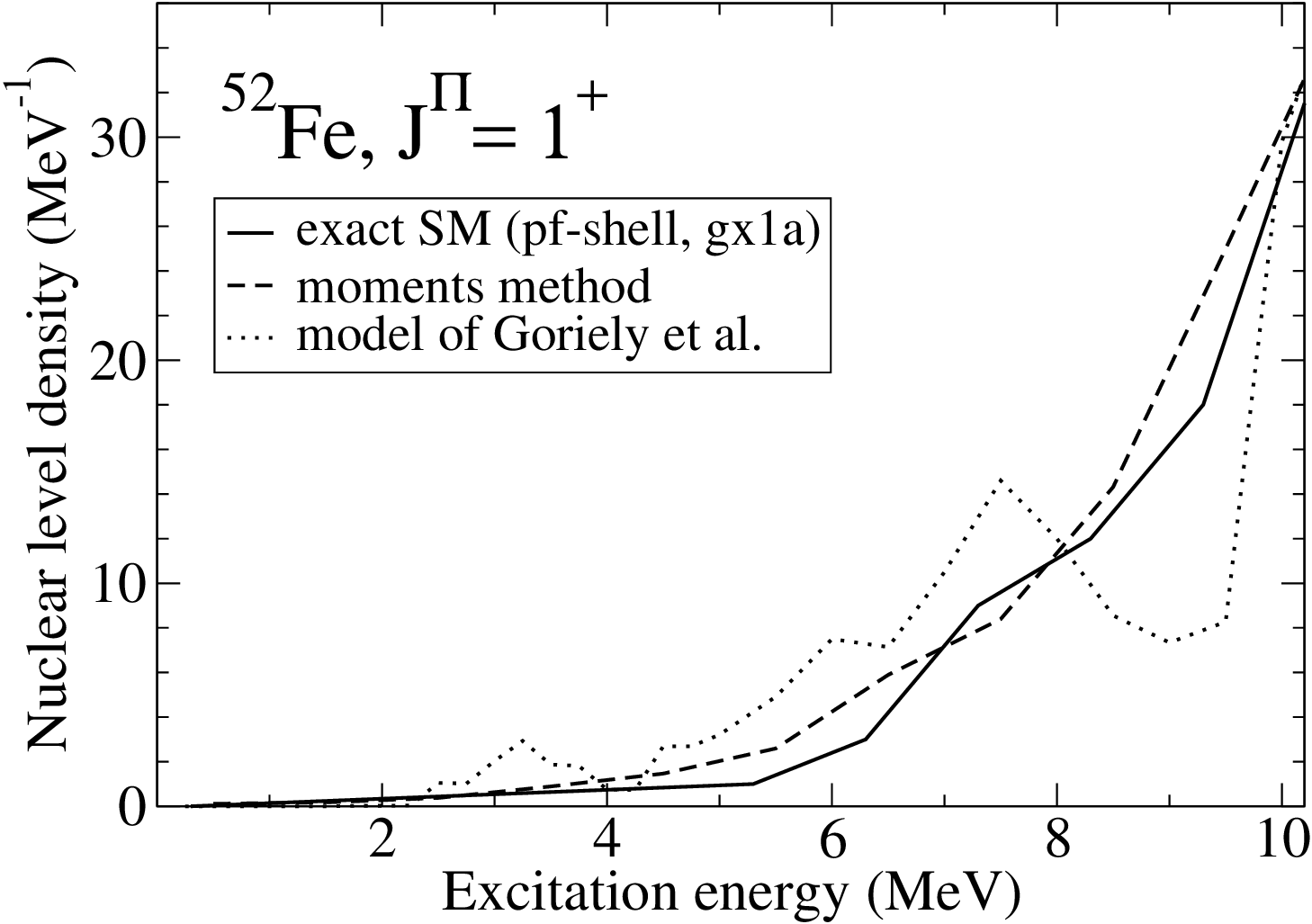}
\vspace*{0.5cm} \\
\includegraphics[width=0.4\textwidth]{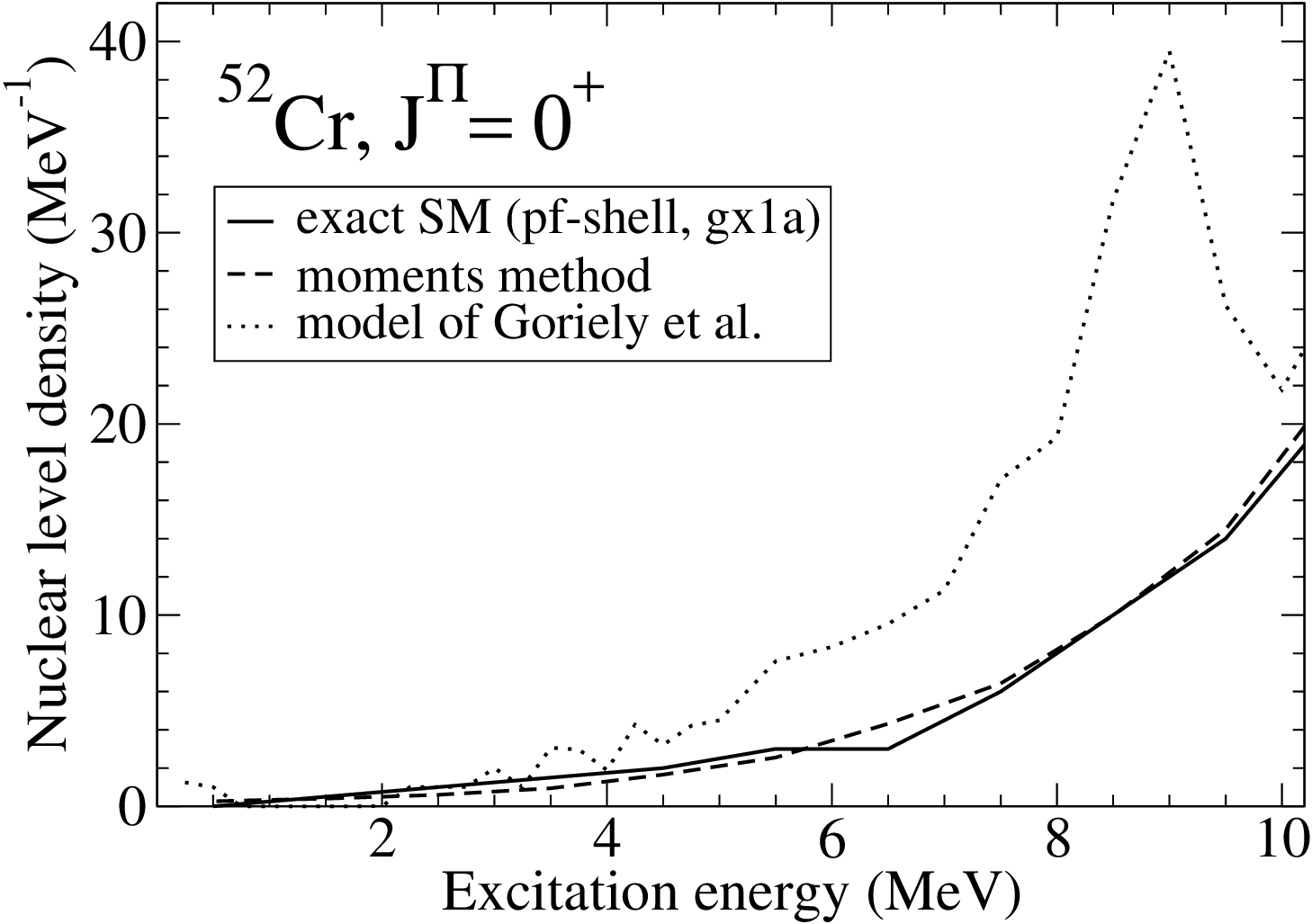} \hspace*{0.5cm} &
\includegraphics[width=0.4\textwidth]{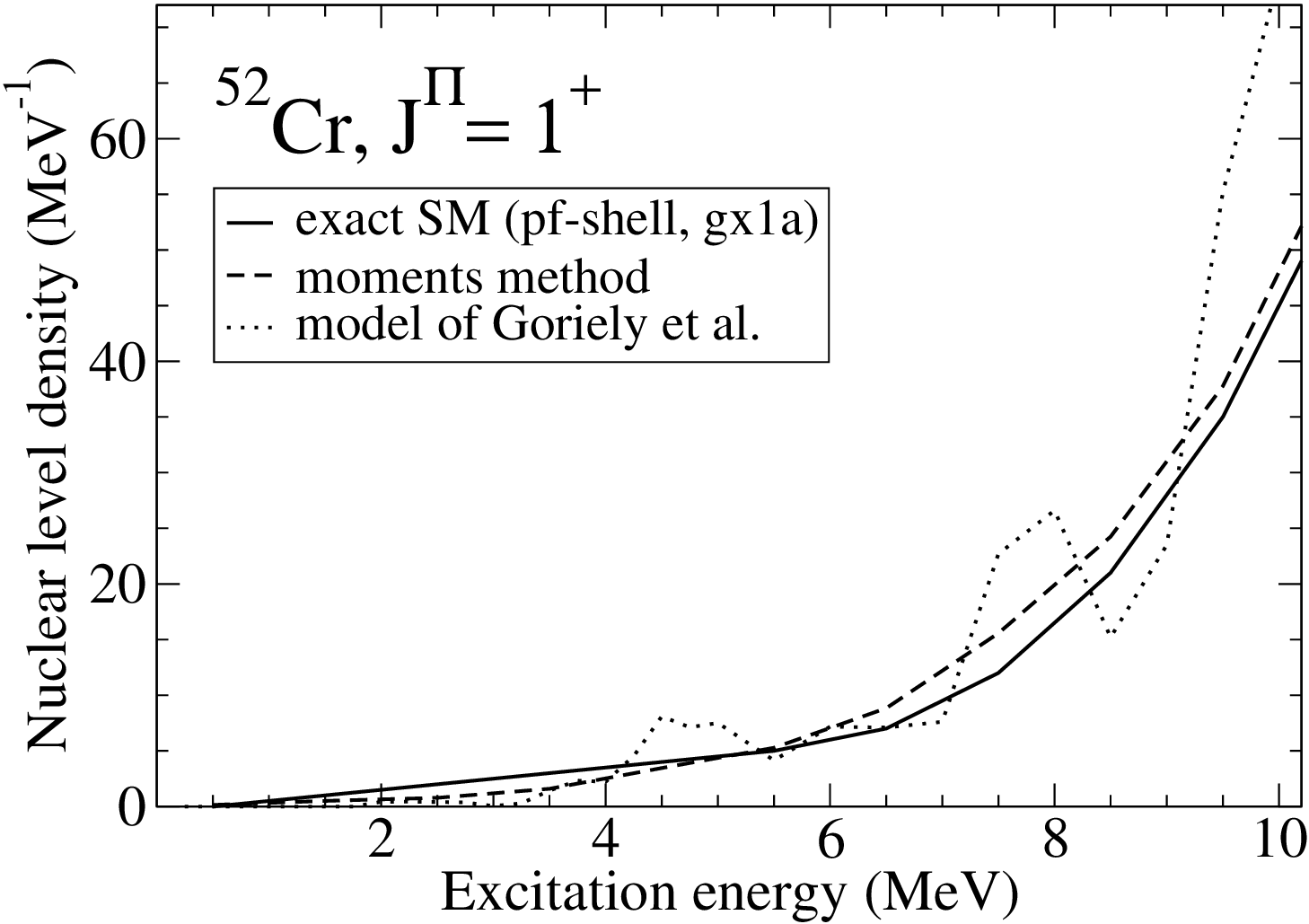}
\end{tabular}
\caption{ Nuclear level densities for $^{52}$Fe and $^{52}$Cr, $\Pi=+1$, different spins: shell model calculation (solid curves) {\sl vs}. moments method (dashed curves) and Hartree-Fock + BCS (dotted corves). Finite-range parameter $\eta=2.8$, $pf$-shell, GXPF1A two-body interaction.\\}
\label{fig6}
\end{figure*}

As stated earlier, the density of low levels with $J^{\Pi}=0^{+}$ in the favorite nucleus $^{28}$Si of the $sd$ shell model is in good agreement with the data. Figs. \ref{fig4} and \ref{fig5} compare the level density calculated with the moments method (solid lines) with the experimental results (stair-cases). The solid stair-cases present an ``optimistic'' attitude when all experimental levels of uncertain parity were counted. Opposite to that, the dashed lines exclude the levels  whose parity is given by the experiment only tentatively. The left graph of Fig. \ref{fig4} shows the summed level density of all positive parity states in $^{28}$Si calculated for the $s+p+sd+pf$ space up to 14 MeV excitation energy. The same quality of comparison can be seen for the positive parity states in $^{26}$Al, the left graph of Fig. \ref{fig5}, the odd-odd nucleus with the relatively well measured energy spectrum.

For the negative parity states, the right parts of  Figs. \ref{fig4} and \ref{fig5}, the shell model predicts more levels than until now have been found experimentally. This can be both due to the imperfection of the shell-model Hamiltonian and because of incompleteness of the data.

\section{Shell-model predictions and mean-field combinatorics}

\begin{figure*}
\begin{tabular}{cc}
\includegraphics[width=0.44\textwidth]{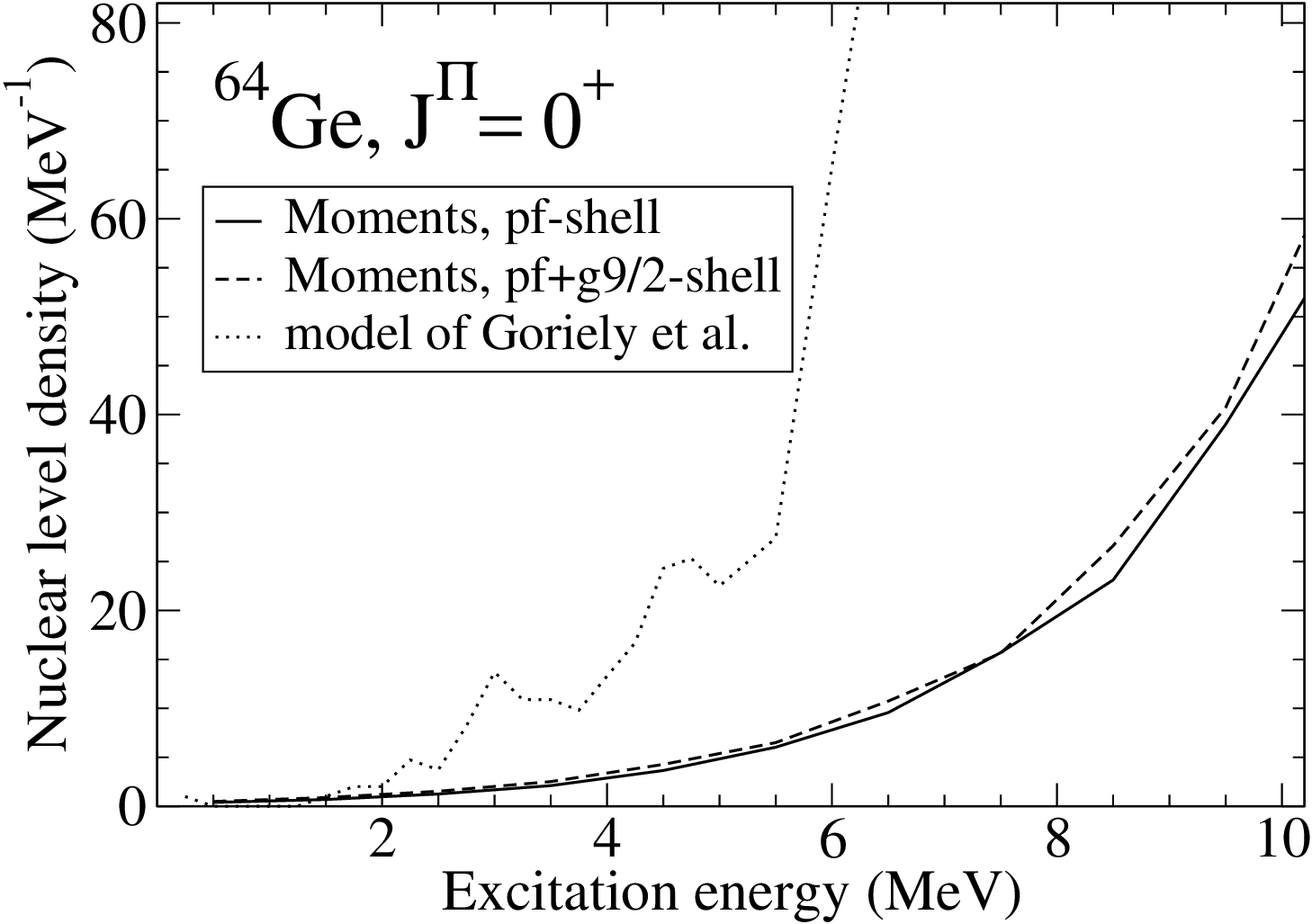} \hspace*{0.5cm} &
\includegraphics[width=0.45\textwidth]{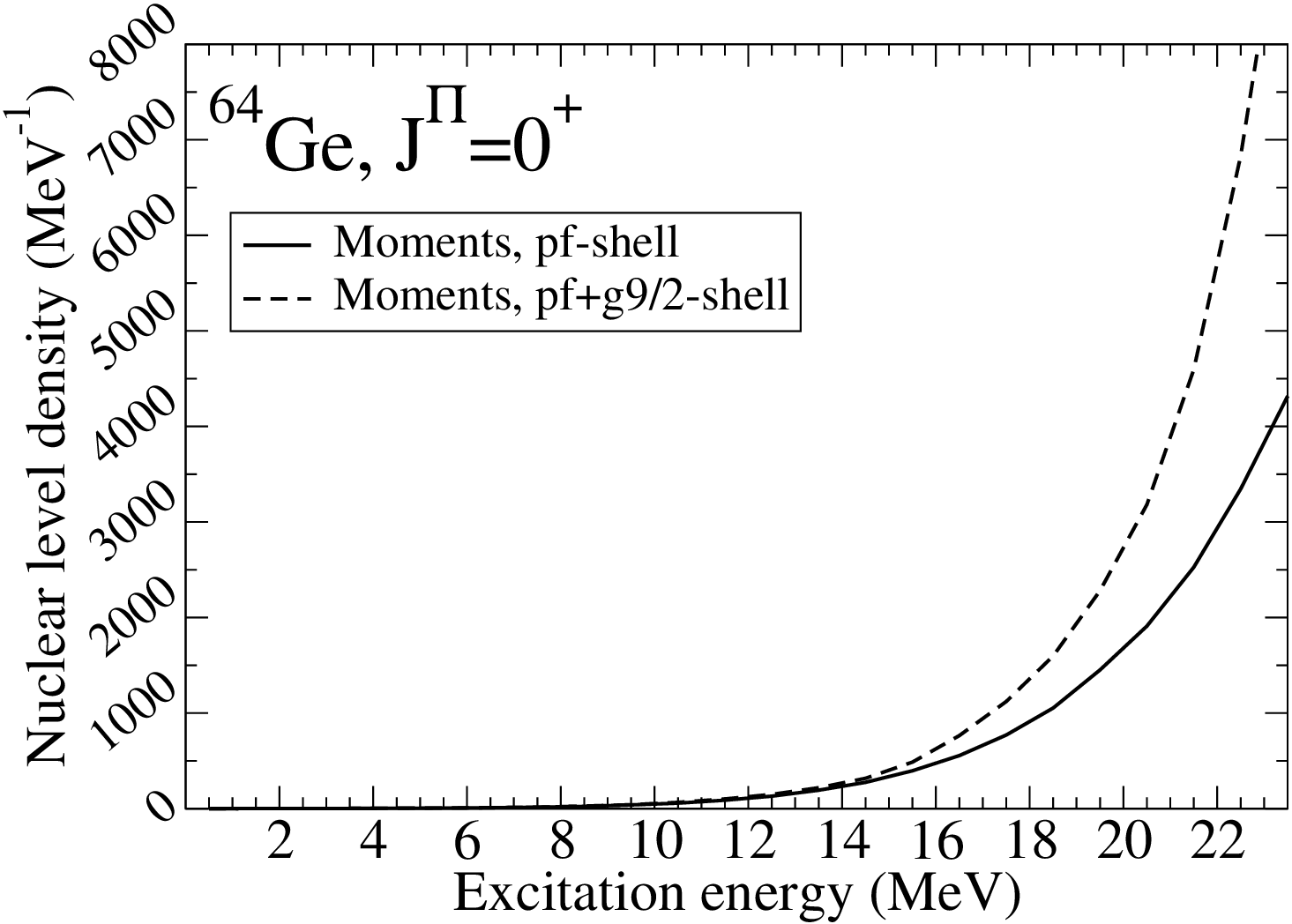}
\end{tabular}
\caption{ Level densities for $^{64}$Ge, spin $J=0$ and $\Pi=+1$. The solid curve presents the calculation in the $pf$ shell with the GXPF1A interaction, the dashed curve corresponds to the calculation in the larger model space with the  level $g_{9/2}$  added, the dotted curve on the left graph presents the results obtained using the HFB single-particle energies and the combinatorial method.\\
\vspace*{0.2cm}}
\label{fig7}
\end{figure*}

\begin{figure*}
\begin{tabular}{cc}
\includegraphics[width=0.45\textwidth]{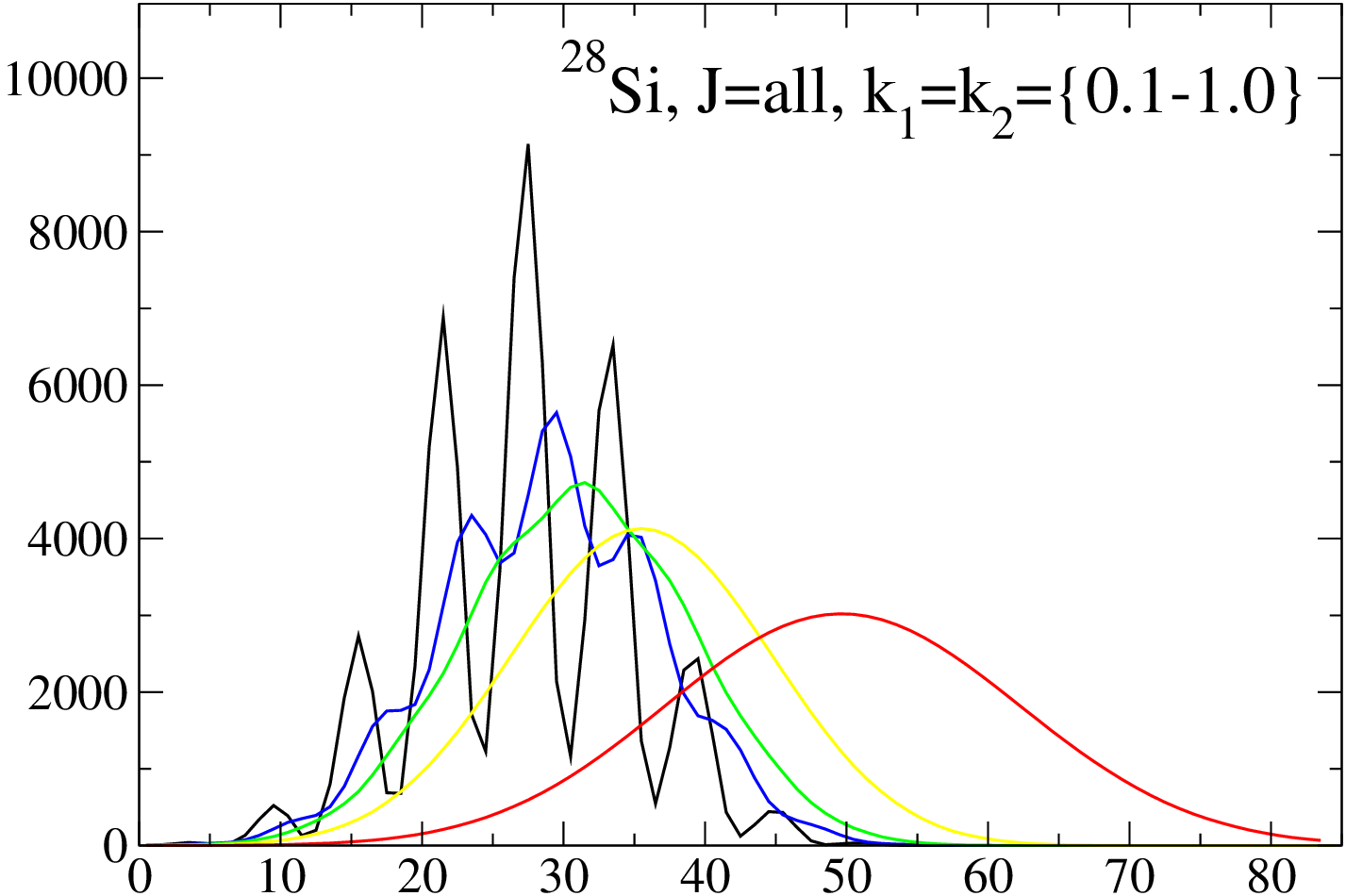} \hspace*{0.5cm} &
\includegraphics[width=0.44\textwidth]{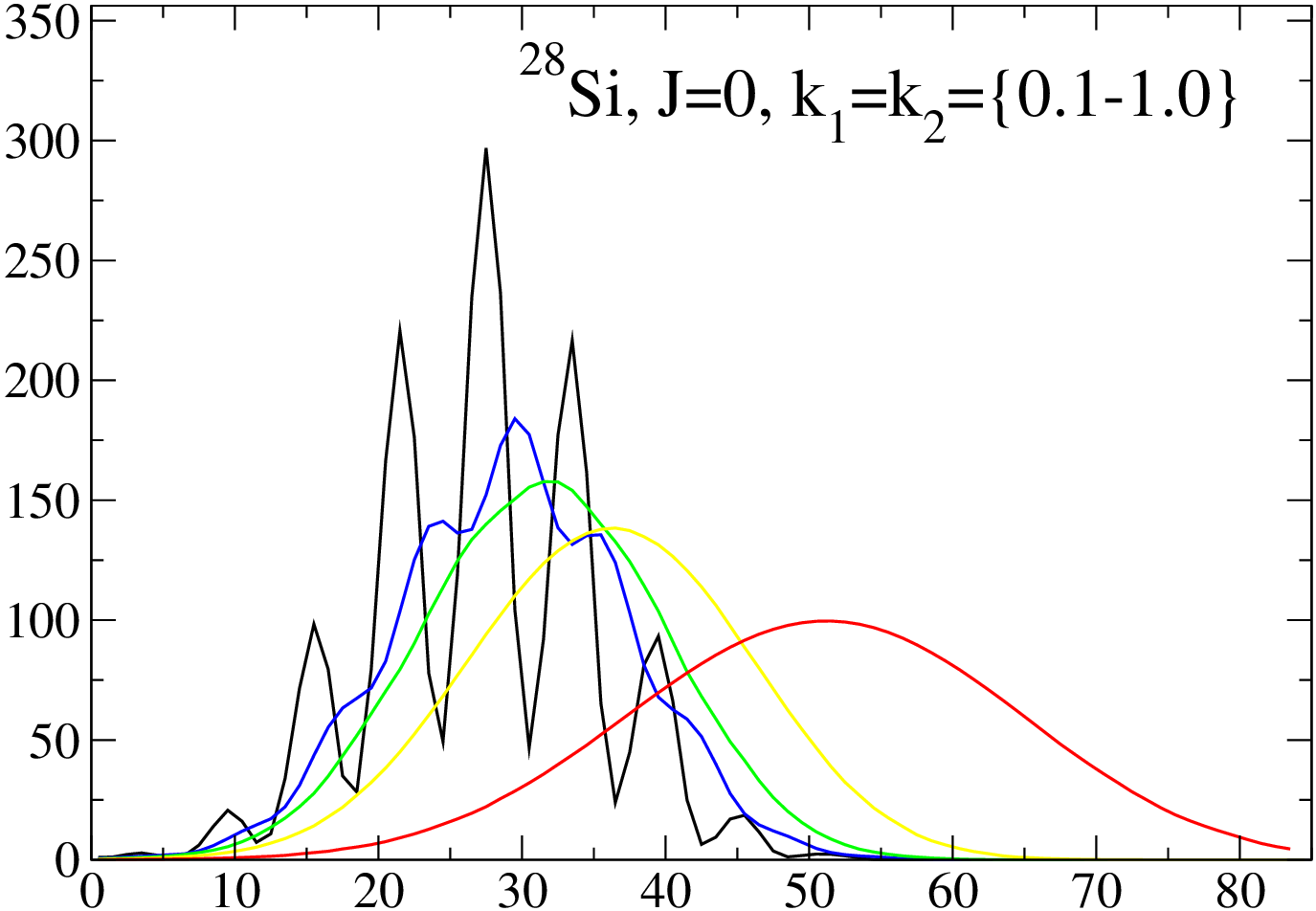}
\end{tabular}
\caption{(Color online.) Level densities for $^{28}$Si, $sd$ model space. Different curves correspond to different scale factors: $k_1=k_2=\{0.1, 0.2, 0.3, 0.5, 1.0\}$ when the pairing and non-pairing parts of the interaction scale similarly. The left graph corresponds to the total density with all $J$ included, while the right graph describes the evolution of the $J=0$ density.\\}
\label{fig8}
\end{figure*}

The widely used standard road to the nuclear level density is going through the mean-field representation of the nuclear dynamics. This traditional approach is based on the classical idea \cite{bethe36,ericson60} of the Fermi-gas where the excited levels result from combinations of many particle-hole excitations. Practically, the combinatorics is used  of single-particle  excitations from the fully occupied Fermi surface identified with the ground state population of the lowest individual orbitals. At low excitation energy, a renormalization of the level density is related to the gap due to the Cooper pairing. In complex nuclei, the low-energy levels observed inside the gap can be interpreted as collective excitations, vibrational or/and rotational. As the collective phenomena of these types correspond typically to the slow self-consistent motion of many particles, it is natural to expect that such coherent combinations of single-particle excitations partly compensate the deficit of levels at low energy due to the pairing gaps and give rise to the so-called {\sl collective enhancement} of the level density \cite{BM,ignatyuk79} in comparison to the single-particle combinatorics of independent particles and holes. Modern refined approaches of this class account in various forms for the pairing phenomenon that changes the excitation spectrum, especially in even-even nuclei \cite{gorieli08,uhrenholt09,hilaire12}.

In the spirit of the mean-field combinatorics, one has to expect the  corresponding suppression of level density at higher excitation energy (damping of collective enhancement); the level density is just redistributed. When the general level density grows, the vibrational modes become strongly mixed with simpler excitations of the two-quasiparticle and more complicated structure, as known very well from the widths of the giant resonances. With smoothing shell gaps, it is harder to distinguish between rotational and intrinsic motion. Recent experiments in nuclei, where low-lying collective excitations are well known, did not find  phenomena of collective enhancement and its fade-out \cite{komarov07}.

\begin{figure*}
\begin{tabular}{cc}
\includegraphics[width=0.45\textwidth]{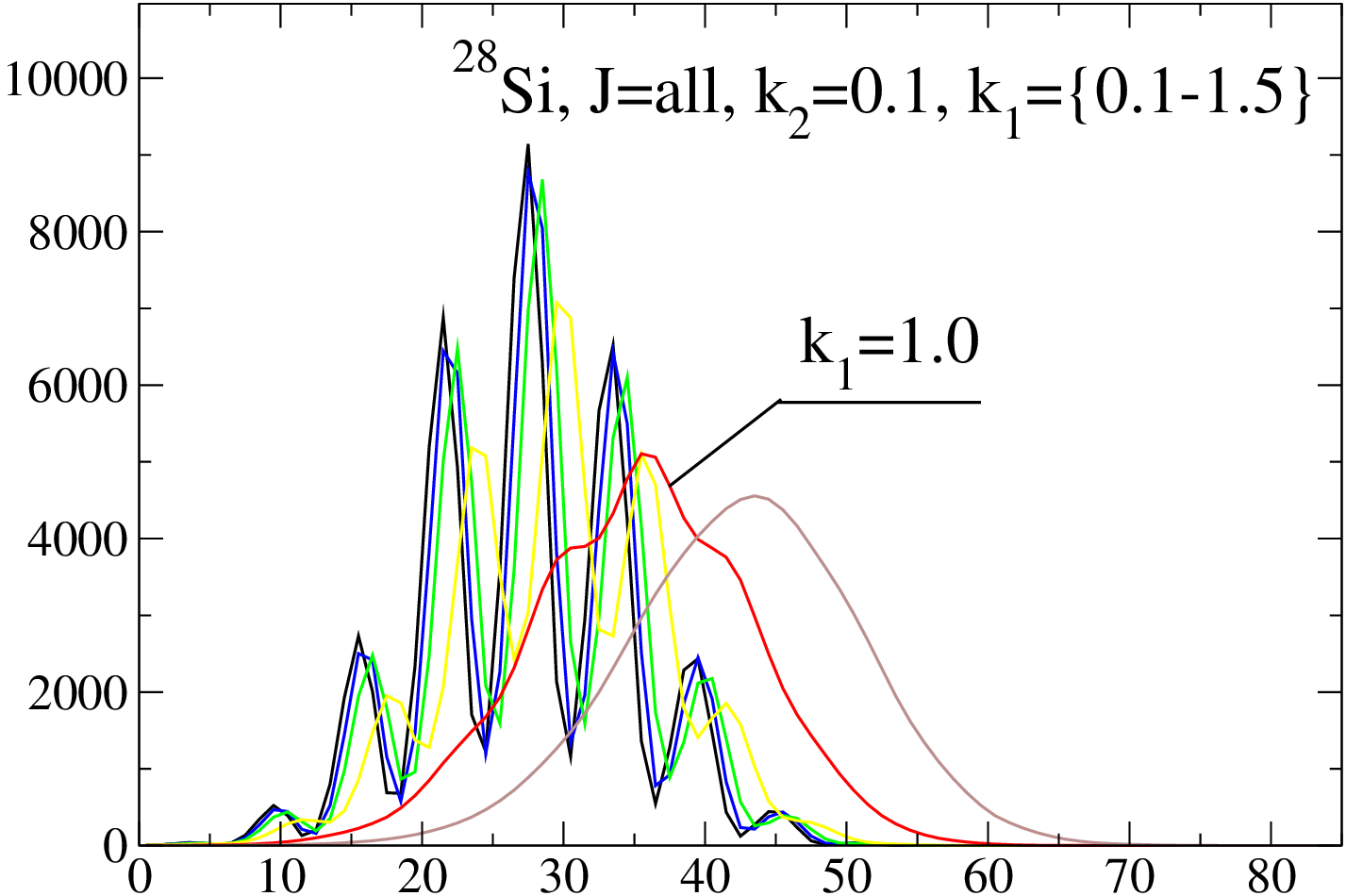} \hspace*{0.5cm} &
\includegraphics[width=0.44\textwidth]{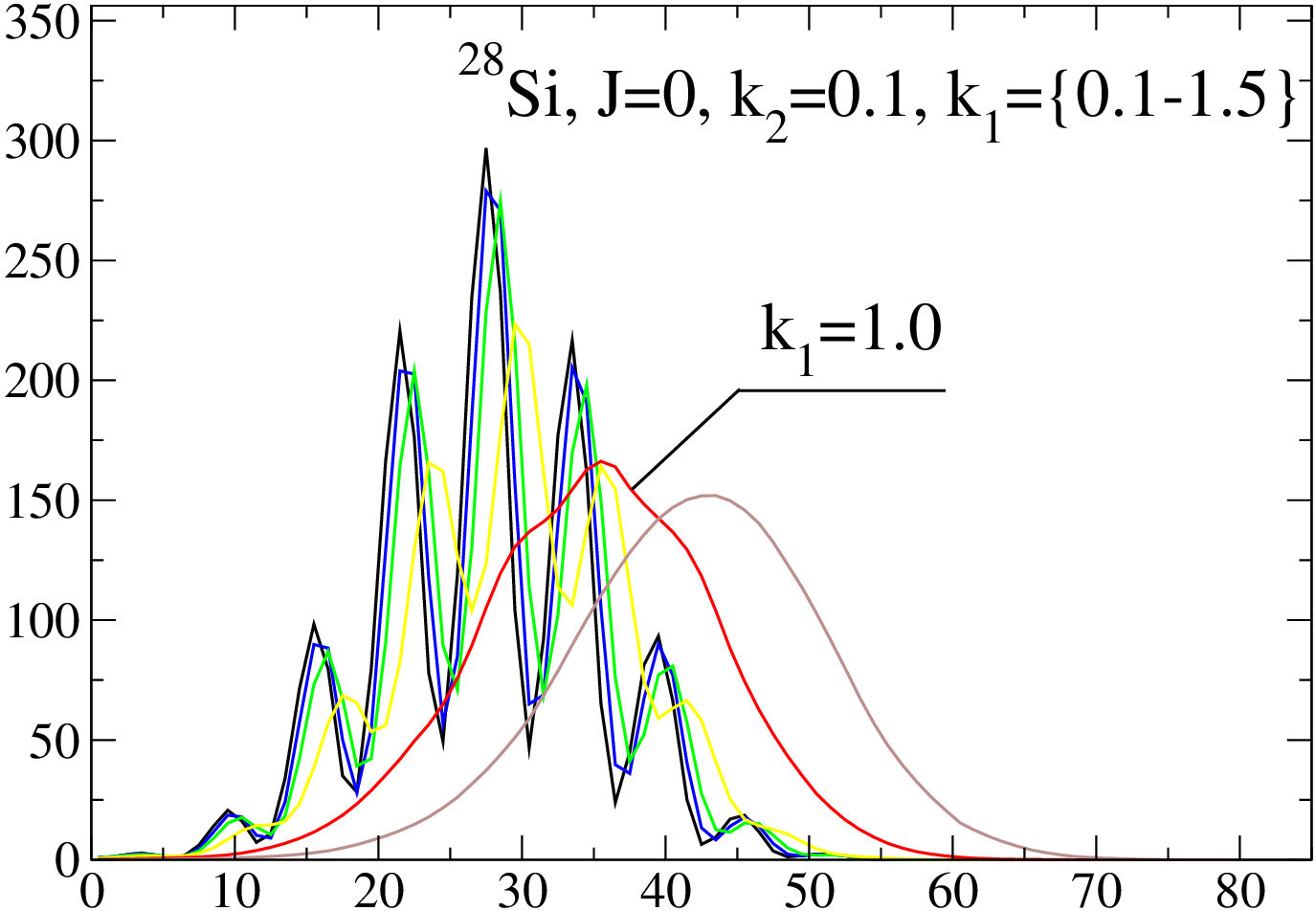}
\end{tabular}
\caption{(Color online.) Level densities for $^{28}$Si, $sd$ model space. The non-pairing interaction is always off: $k_2=0.1$, while the pairing interaction scales, $k_1=\{0.1, 0.2, 0.3, 0.5, 1.0, 1.5\}$. The left graph corresponds to the total density with all $J$ included, while the right graph describes the evolution of the $J=0$ density.\\
\vspace*{0.3cm}}
\label{fig9}
\end{figure*}

\begin{figure*}
\centering
\begin{minipage}[t]{.44\textwidth}
\includegraphics[width=1.00\linewidth]{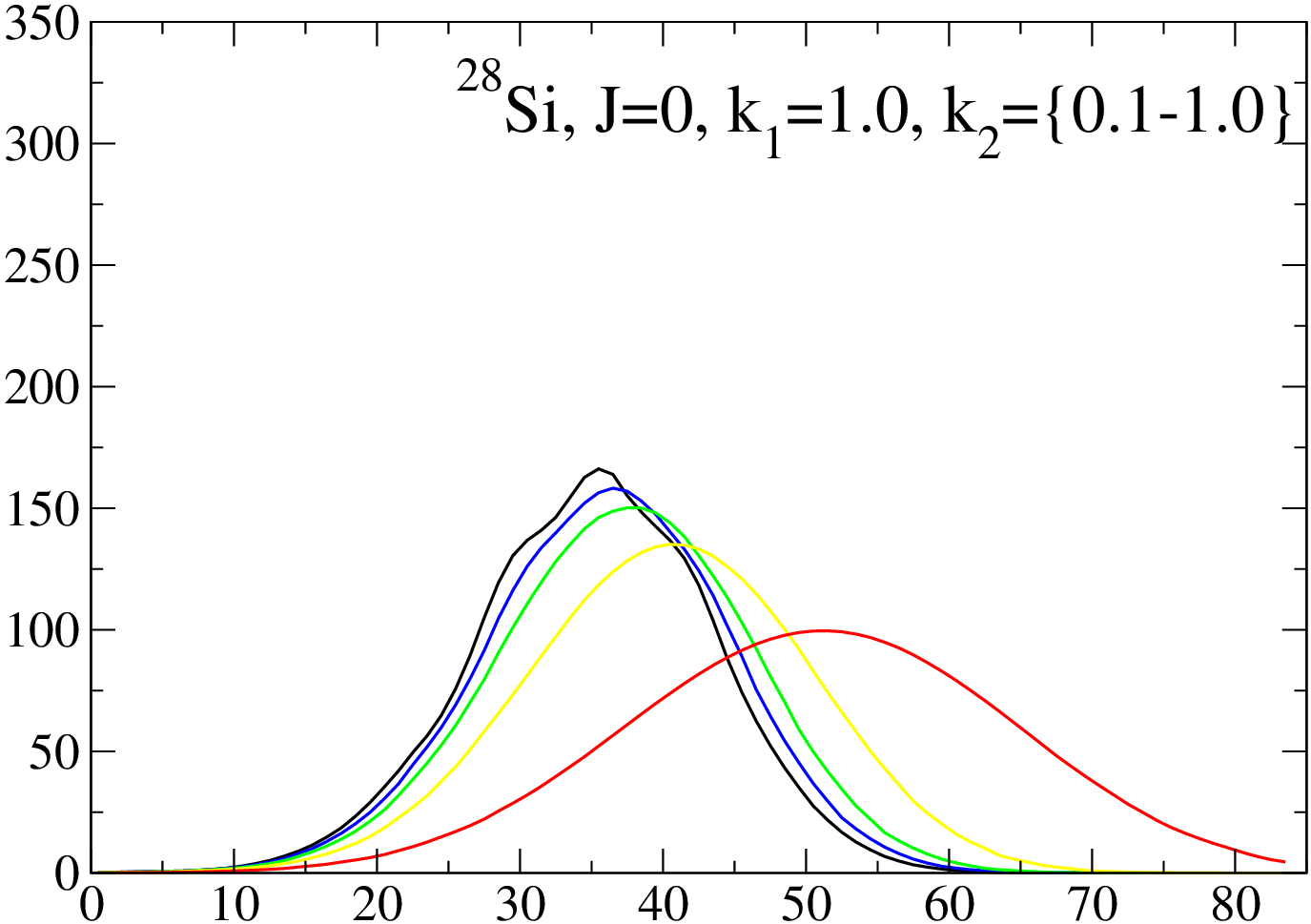}
\caption{(Color online.)  Level densities for $^{28}$Si, $J=0$, $sd$ model space. The black curve presents $k_1=1.0, k_2=0.1$, then the remaining parts of the interaction are increased together with $k_2$ up to the red curve ($k_1=k_2=1.0$) that shows the density for the realistic interaction.}\label{fig10}
\end{minipage}\qquad
\begin{minipage}[t]{.46\textwidth}
\includegraphics[width=1.00\linewidth]{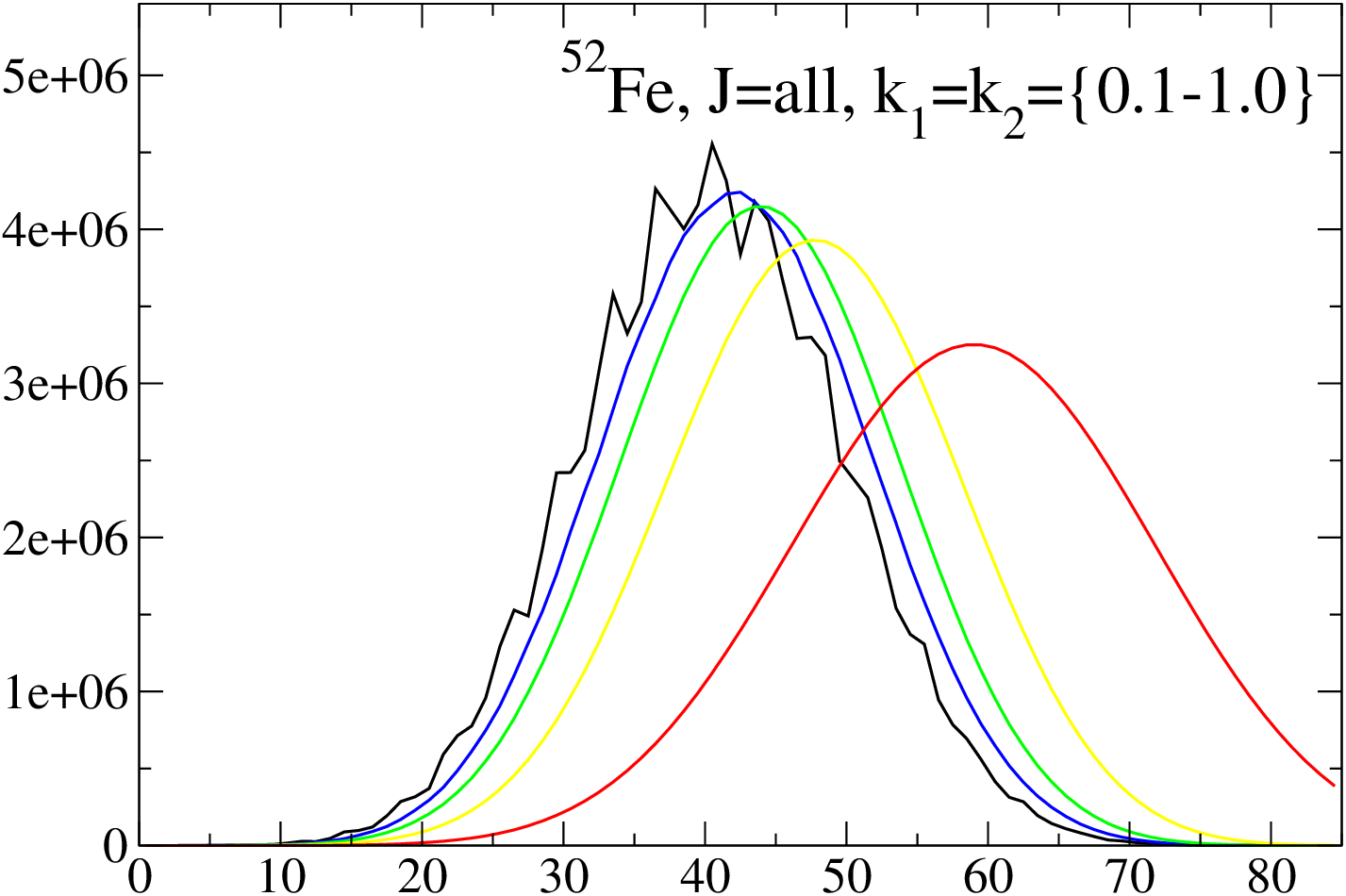}
\caption{(Color online.)  Level densities for $^{52}$Fe, all $J$, $pf$ model space. This figure and its color scheme are similar to the left panel on Fig. \ref{fig8}.}\label{fig11}
\end{minipage}
\end{figure*}

In not too heavy nuclei, the quasiparticle combinatorics (on the base of the BCS or Hartree-Fock-Bogoliubov pairing description) reveals step-wise effects of subshell occupation and pair breaking. This leads, at relatively low energy, to the irregular picture of the level density that clearly reflects these steps. The shell-model Hamiltonians, as a rule, contain all interaction matrix elements allowed by the selection rules. One of the main conclusions of the full shell-model calculation is that the presence of all interactions is significantly smoothing the whole picture so that it is hard to see the traces of individual families which could be still recognized only by the special observables and selection rules for the individual transitions. We can recall that our algorithm still starts with the partitions
formed by independent particles which then overlap and lose their boundaries.

The shell-model Hamiltonian contains all pairing matrix elements (and not in the simplified form with constant matrix elements) as well as the interaction processes responsible for multipole-multipole forces and deformation. Therefore all collective effects mentioned above are fully taken into account if the orbital space is sufficiently broad. Fig. \ref{fig6} shows the comparison of the level densities for the states $0^{+}$ and $1^{+}$ in the nuclei $^{52}$Fe and $^{52}$Cr of the $pf$-shell. The thin dotted lines give the level density found with the mean-field combinatorics built on the Hartree-Fock mean field and BCS pairing description. All irregularities of the level density found through the mean-field combinatorics are completely smoothed in the full moments calculation. This is a typical result encountered in all examples. Again we see that the method under discussion produces the level density practically identical to the full shell-model diagonalization when the latter is possible.

Fig. \ref{fig7} illustrates the influence of the enlargement of the orbital shell-model space. The level density of states $0^{+}$ in the $N=Z$ nucleus $^{64}$Ge becomes sensitive to the inclusion of the next shell ($g_{9/2}$) only at the excitation energy greater than 14 MeV (see the right graph of Fig. \ref{fig7}) which means that the region of neutron resonances could be reliably evaluated with the more narrow orbital space. This case has important ramifications for the astrophysical consideration of the element abundance since this nucleus is considered to be a waiting point in the $r$-process
of nucleosynthesis.

\section{Coherent and incoherent interactions}

\begin{figure*}
\begin{tabular}{cc}
\includegraphics[width=0.45\textwidth]{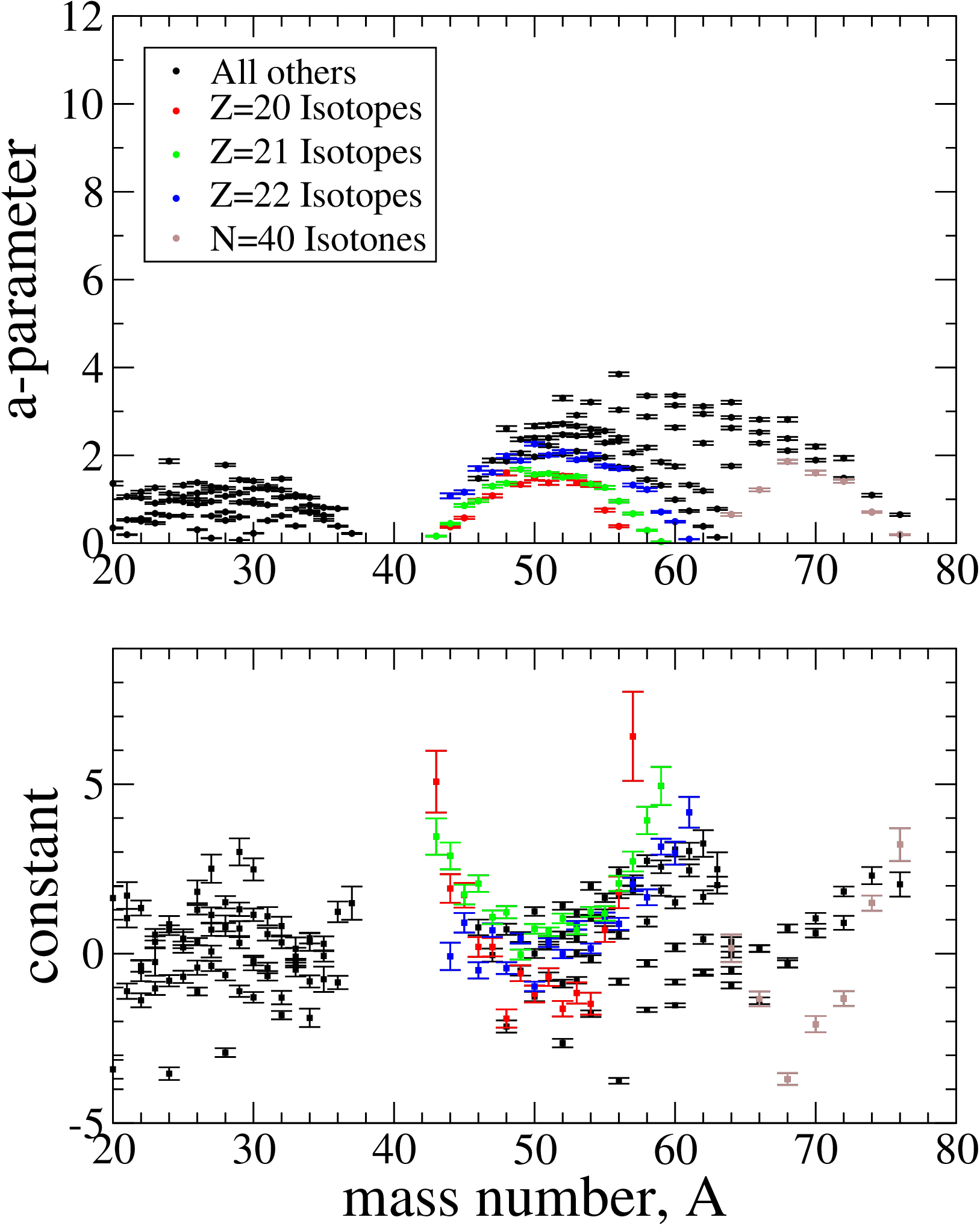} \hspace*{0.5cm} &
\includegraphics[width=0.45\textwidth]{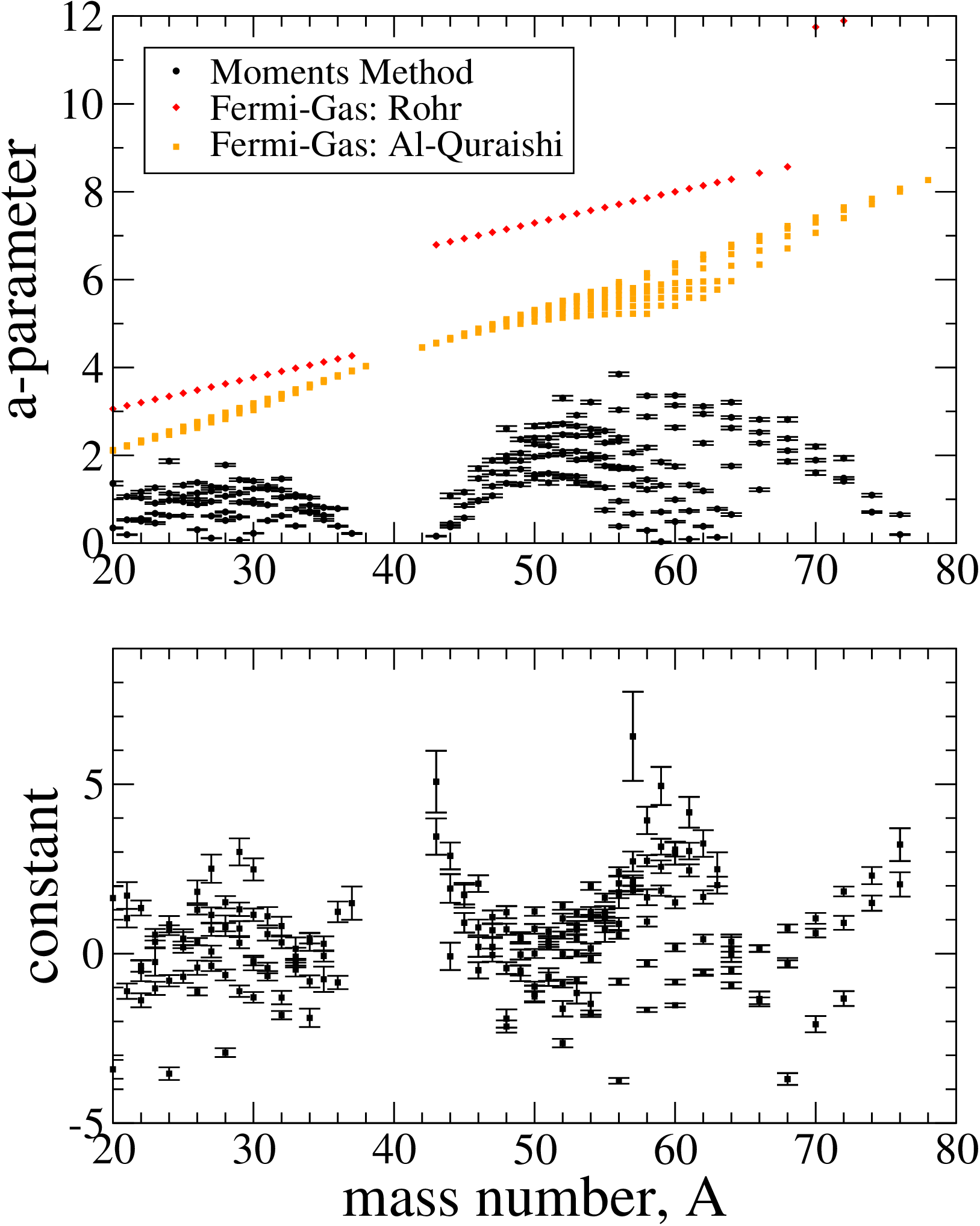}
\end{tabular}
\caption{(Color online.) Interpolation of the single-particle level density parameter $a$,  Eq. (\ref{15}). Left panel: different colors present different isotopes or isotones. Right panel: moments-method calculation with interpolation (black circles), fit using the  experimental data on neutron resonances (Ref. \cite{rohr84}, orange diamonds), and fit using experimental low-lying levels (\cite{alquraishi01}, yellow squares).\\}
\label{fig12}
\end{figure*}

As the whole shell-model Hamiltonian contributes to the traces defining the level density, we can explore the effects of individual components of the effective interactions including the ``incoherent'' parts of the full Hamiltonian which do not significantly contribute to the formation of the mean field. These parts of the interaction determine the finite lifetime of the simple quasiparticle (or collective) modes and their fragmentation in terms of genuine complicated eigenstates of exceedingly entangled nature. In particular, these collision-like interactions are responsible for the formation of chaotic states with high information entropy and the process of thermalization \cite{big}. For example, it was shown \cite{ZMV01,VZB02,AARZ12} that the exactly considered pairing interaction contains some chaotic features but they are still not sufficient for establishing the complete chaotic picture comparable to the predictions of the Gaussian Orthogonal Ensemble.

Below we show the evolution of the level density as a function of the interaction modes included in the full calculation of the moments. Although it is not difficult to vary all individual matrix elements, here we use a simplified approach, presenting the whole two-body interaction Hamiltonian in the $sd$-shell model as consisting of two parts with variable intensity,
\begin{equation}
H=h+k_{1}V({\rm pairing})+k_{2}V({\rm non-pairing}).                                        \label{9}
\end{equation}
Here $h$ contains the single-particle energies, $V({\rm pairing})$ all matrix elements with the pairs of nucleons in the channel $J^{\pi}T=0^{+}1$, while all other shell-model matrix elements are attributed to the last term. The numerical coefficients $k_{1}$ and $k_{2}$ are varied giving rise to different versions of the shell model; the realistic case emerges at $k_{1}=k_{2}=1$. It is easy to understand that in the moments method (and in the full diagonalization, see \cite{big}) new independent components of the Hamiltonian add in quadratures to the final width of the shell-model level density.

\begin{figure*}
\centering
\begin{minipage}[t]{.48\textwidth}
\includegraphics[width=1.00\linewidth]{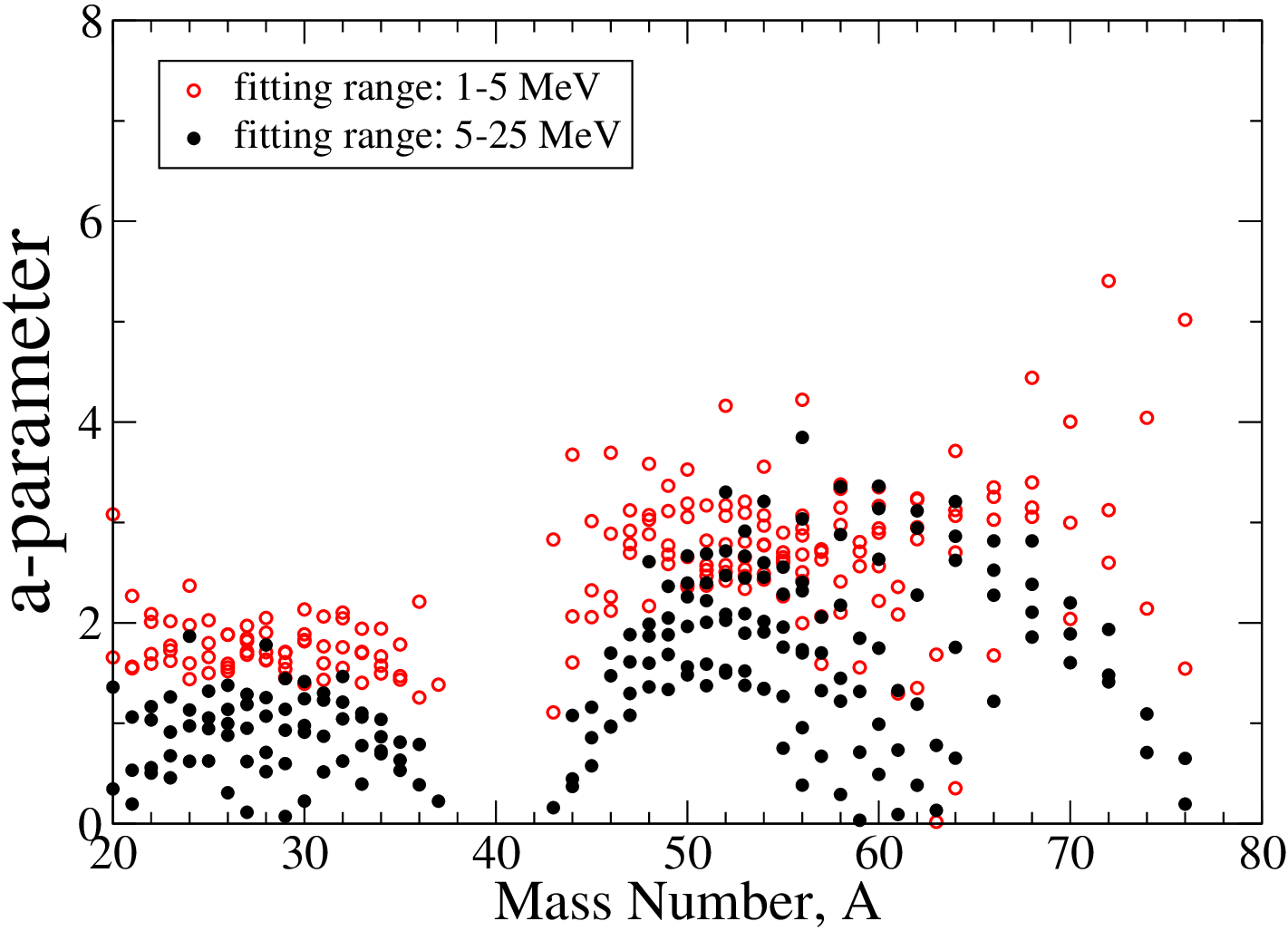}
\caption{(Color online.) Level density parameter (\ref{11}) fitted in the simple Fermi-gas model for $sd$ and $pf$ shell nuclei. The empty circles (red color) present the fitting range 1-5 MeV, the filled black circles correspond to the  fitting range 5-25 MeV.}\label{fig13}
\end{minipage}\qquad
\begin{minipage}[t]{.48\textwidth}
\includegraphics[width=1.00\linewidth]{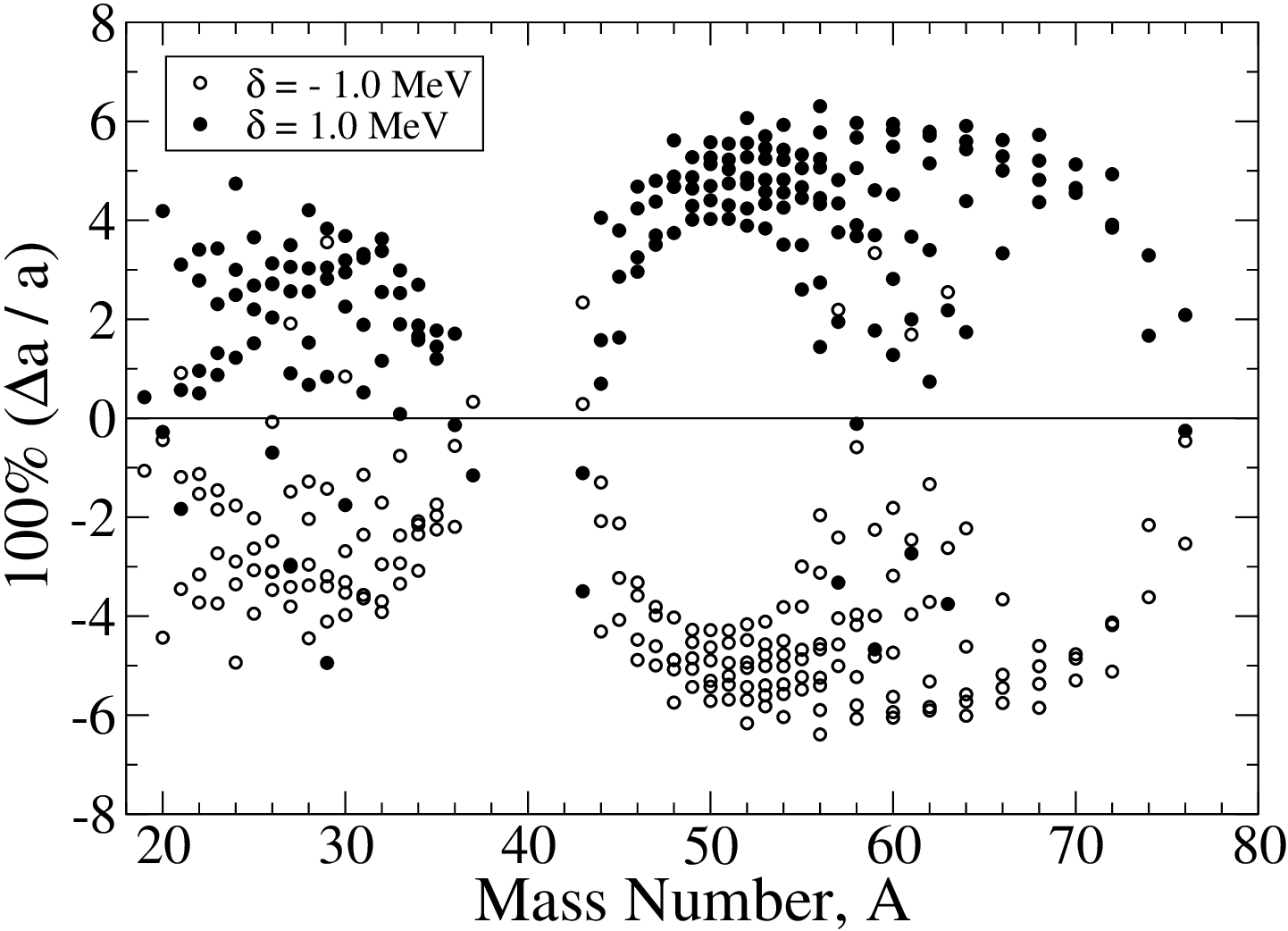}
\caption{ The relative changes in the level density parameter $a$ due to the pairing energy shift $\delta$, Eq. (\ref{17}), shown for $sd$- and
$pf$-shell nuclei. The empty circles present the fit with the energy shift $\delta=-1$ MeV and the filled circles correspond to $\delta=1$ MeV. \\}\label{fig14}
\end{minipage}
\vspace*{0.2cm}
\end{figure*}

The global evolution of the full level density in $^{28}$Si as the coefficients $k_{1}=k_{2}$ are varied from 0.1 to their realistic values is shown in the left graph of Fig. \ref{fig8}. The low residual interaction obviously keeps untouched the independent-particle partition structure of the Hilbert space that reminds the results of the mean-field combinatorics. As the parameters $k_{1},k_{2}$ grow, the next curves show the development of the final picture. The configurational structure is gradually washed out by the residual interaction leading to the final smooth level density discussed earlier. Let us stress that the observed evolution is not a consequence of the superposition of all subspaces with different values of $J$. The individual subspace $J^{\Pi}=0^{+}$, the right graph on Fig. \ref{fig8}, demonstrates practically the same evolution.

Fig. \ref{fig9} describes the situation when the non-pairing components of interaction are suppressed, $k_{2}=0.1$,  but the pairing strength evolves. Superposing all values of $J$ (the left graph of Fig. \ref{fig9}), we see that the smooth Gaussian-like curve is achieved only at the non-realistically high pairing strength. Again the picture is nearly the same when only the states $J=0$ are considered, the right graph of Fig. \ref{fig9}. The realistic pairing strength, $k_{1}=1.0$, at the absence of non-pairing interactions is still not sufficient for the fully smooth level density. At low excitation energy $<20$ MeV the evolution of the level density for $J=0$ clearly shows the disappearance of the typical large oscillations with the growth of pairing. Here, indeed, the pairing interaction shifts the noticeable part of levels to higher energies. If the pairing matrix elements are fixed at the empirical value, Fig. \ref{fig10}, the large bumps from the original partitions do not appear but the incoherent interactions very much broaden the final result. The generic character of this scenario is confirmed by  Fig. \ref{fig11} for $^{52}$Fe.

\section{Thermodynamic description and comparison with phenomenology}

\subsection{Simple Fermi-gas}

\begin{figure*}
\centering
\begin{minipage}[t]{.48\textwidth}
\includegraphics[width=1.00\linewidth]{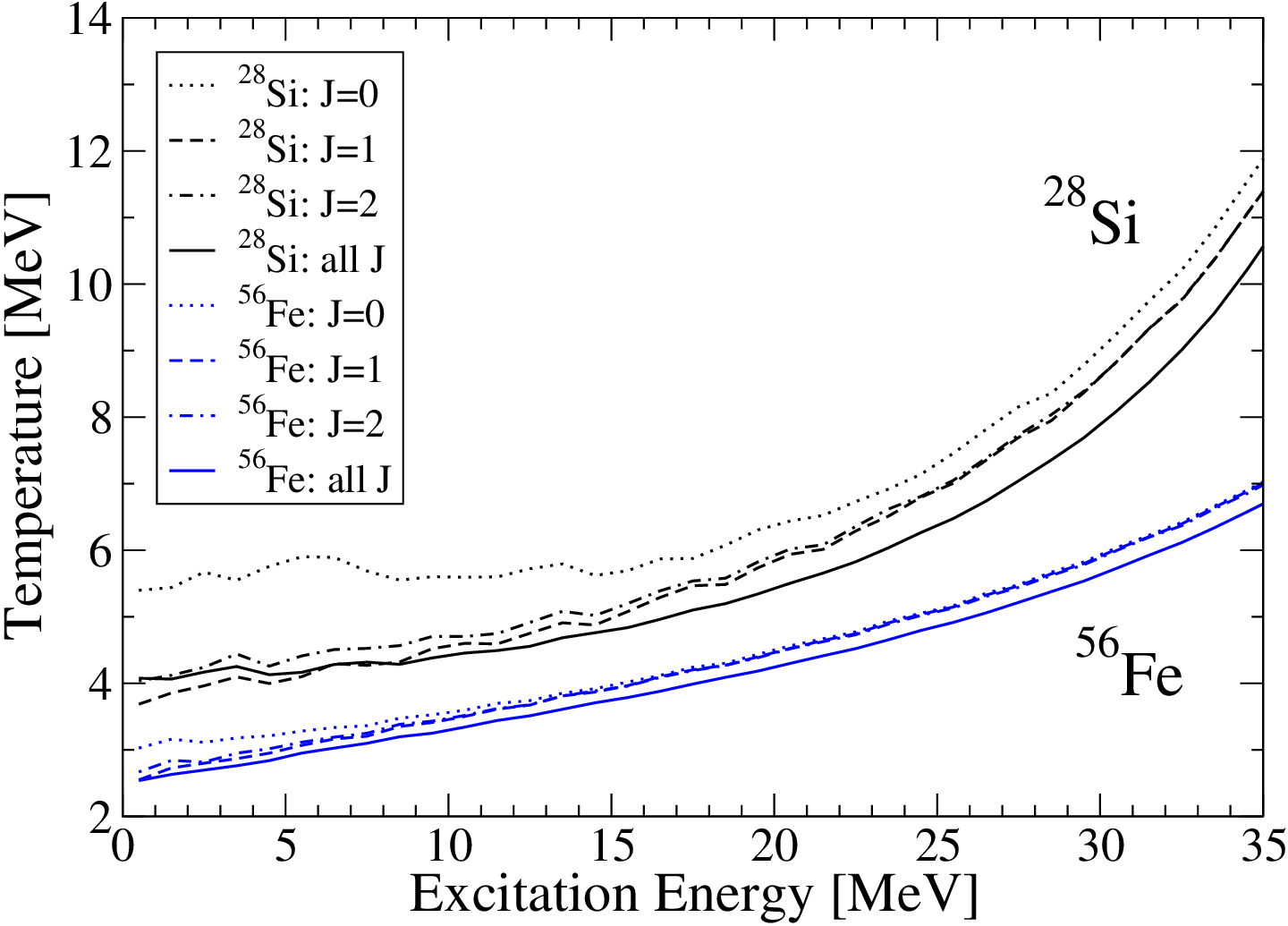}
\caption{(Color online.) Thermodynamic temperature as a function of excitation energy, Eq. (\ref{18}), calculated in the moments method. The top four curves (black color) correspond to the densities of $^{28}$Si and the bottom four curves (blue color) correspond to the  densities of $^{56}$Fe. }\label{fig15}
\end{minipage}\qquad
\begin{minipage}[t]{.48\textwidth}
\includegraphics[width=1.00\linewidth]{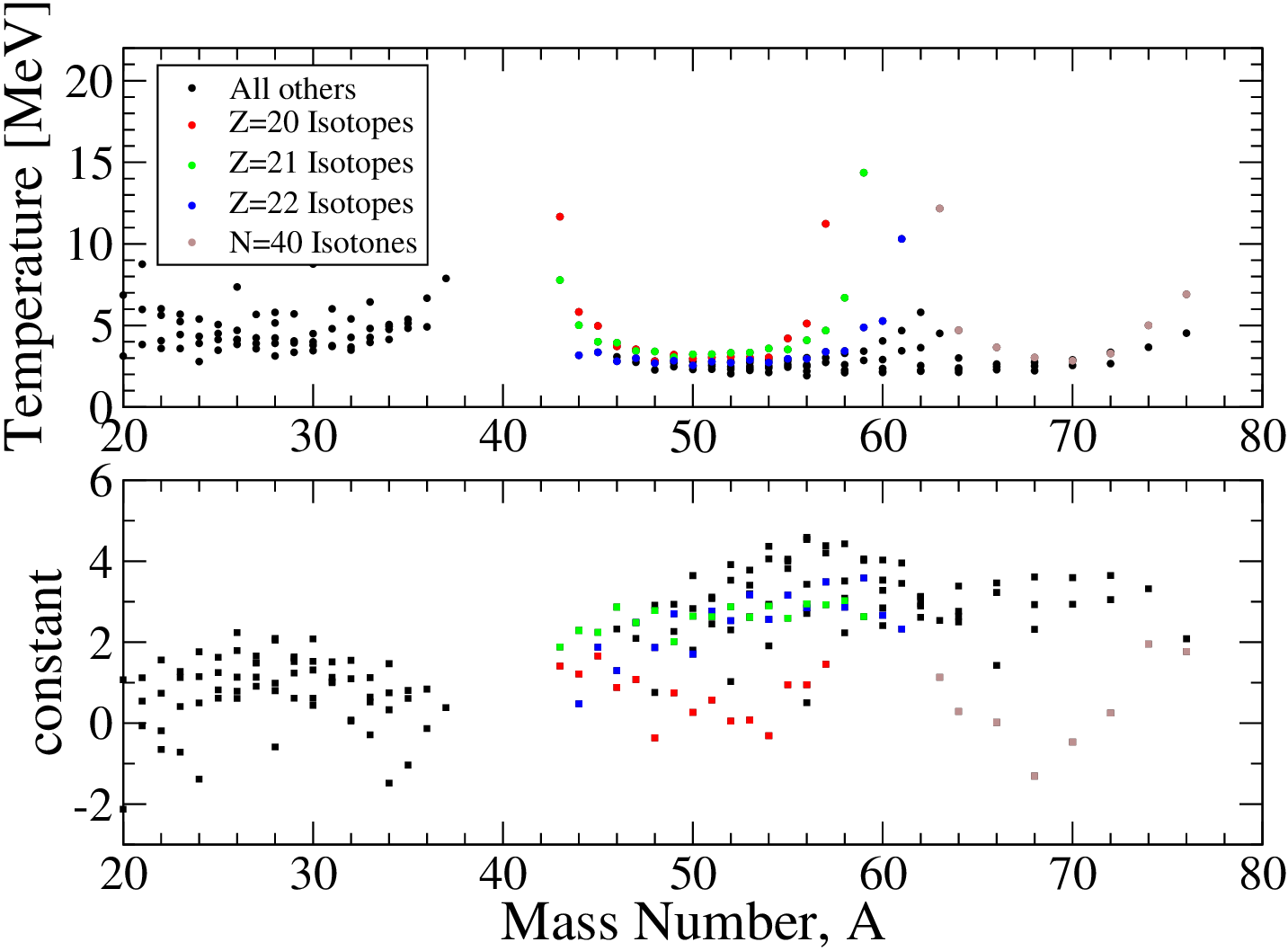}
\caption{(Color online.) The constant temperature $T$ and the ``constant" (see Eq. (\ref{19})) fitted for the  energy range 5-15 MeV. Different sets of nuclei from the $pf$ shell with complete and almost complete shells are presented by different colors: $Z=20$ (red), $Z=21$ (green), $Z=22$ (blue), and $N=40$ (brown).\\
\vspace*{0.2cm}}\label{fig16}
\end{minipage}
\end{figure*}

The shell-model Hamiltonian (\ref{1}) starts from  non-interacting particles or quasiparticles, elementary excitations in the mean field of certain symmetry that determines the appropriate quantum numbers of excited states. For nuclei, the adequate image is that of the perfect two-component Fermi gas. The ground state of the system is the filled Fermi sphere, and the excited states are described by the particle-hole picture. In the realistic many-body physics, this is just an initial step that has to be followed by switching on the interaction between particles and holes. However, already by this mechanism, the level density increases exponentially which justifies the traditional phenomenological approaches.

The particle-hole phenomenology uses the steepest descent method to calculate the level density as a function of excitation energy $E$ through the Laplace transform of the partition function which leads to the standard result for one type of particles,
\begin{equation}
\rho(E)=\,\frac{1}{4\sqrt{3}\,E}\,e^{2\sqrt{aE}},                                         \label{10}
\end{equation}
where the {\sl level density parameter} is
\begin{equation}
a=\,\frac{\pi^{2}}{6}\,\nu_{F}                                                                  \label{11}
\end{equation}
and $\nu_{F}$ is the density of {\sl single-particle} states at the Fermi surface. The generalization to a proton-neutron system leads to a modified expression,
\begin{equation}
\begin{aligned}
\rho(E) =\,\frac{\sqrt{\pi}\,\bar{a}}{12\,(aE)^{5/4}}\,e^{2\sqrt{aE}}\\
 =\,\frac{6^{1/4}\,\bar{\nu}}{12\,(\nu_{F}E)^{5/4}}\,e^{2\sqrt{aE}},
\end{aligned}\label{12}
\end{equation}
where the parameters $a$ and $\bar{a}$ now include the total single-particle density of states at the Fermi surface, $\nu_{F}=\nu_{F}(n)+\nu_{F}(p)$, and the effective single-particle density $\bar{\nu}=\nu_{F}^{2}/(2\sqrt{\nu_{F}(n)\nu_{F}(p)}),$ correspondingly, see Eq. (\ref{11}). The singularities at $E\rightarrow 0$ in densities (\ref{10}) and (\ref{12}) show that the statistical method of calculation is invalid at too low excitation energy where the number of nuclear many-body states is small. The notion of the level density requires that the excitation energy be greater than the average distance between the single-particle levels,
\begin{equation}
E \gg \frac{1}{\nu_F}. \label{13}
\end{equation}
In thermodynamic language, the nuclear temperature $t$ is introduced through the Fermi-gas formula for the excitation energy,
\begin{equation}
E=at^{2}.                                                                                                    \label{14}
\end{equation}
In general, for the low excitation energy region it is always recommended to use directly the available experimental information.

If one tries to compare the thermodynamic level density of the Fermi-gas type with experimental data, it is hard to expect the numerical agreement of the level density parameter (\ref{11}) with that required by data even if the exponential growth of the total level density takes place. As mentioned in the Introduction, when the level density grows, the residual interactions lead to multiple avoided crossings and mixing of many-body levels with the same exact quantum numbers. This process of chaotization evolves the level network considered as a function of the interaction strength close to the aperiodic crystal with a small average spacing. The whole set of stationary states is becoming locally close to the predictions of the Gaussian Orthogonal Ensemble of random matrices. The wave functions here are quite complicated superpositions of very many particle-hole states. The energy behavior of the level density is now close to a Gaussian \cite{big} with the total width that is given by adding in quadratures the initial width due to the mean-field quasiparticles and the dispersion of the off-diagonal matrix elements of residual interactions.

Fig. \ref{fig12} shows the parameters $a$ (the top panels) and $constant$ (the bottom panels) found from the  shell-model calculation of the total level density fitted by
\begin{equation}
\ln[\rho(E,M=0)]=2\sqrt{aE}\,-\,\frac{5}{4}\,\ln E+{\rm const}.                 \label{15}
\end{equation}
Both the parameter $a$ and the constant show the change clearly correlated with the microscopic filling of the nuclear shells. The parameter $a$ reveals the maximum in the middle of the shell occupation as it should have been expected from the construction of the model. The empirical estimates for the same nuclei are available from the level density at the neutron resonances energy \cite{rohr84}, which do not show considerable shell effects, and by extrapolation from low-lying levels \cite{alquraishi01} where one can see very weak shell effects in the region of the mass number around $A\approx 50$. The constant in Eq. (\ref{15}) is small but also shows in some cases the shell-model dependence with a minimum in counter-phase with the parameter $a$.
Fig. \ref{fig13} shows the dependence of the level density parameter $a$ on the energy range for which the fitting was performed. The empty circles (red color) present the low energy fit, around 1-5 MeV, while the filled circles (black color) present the standard 5-25 MeV energy range fit. We can see that at low energies the fitted level density parameter $a$ is slightly larger, but still it is not large enough to be compared with the empirical estimates \cite{rohr84,alquraishi01}.

\begin{figure*}
\centering
\begin{minipage}[t]{.48\textwidth}
\includegraphics[width=1.00\linewidth]{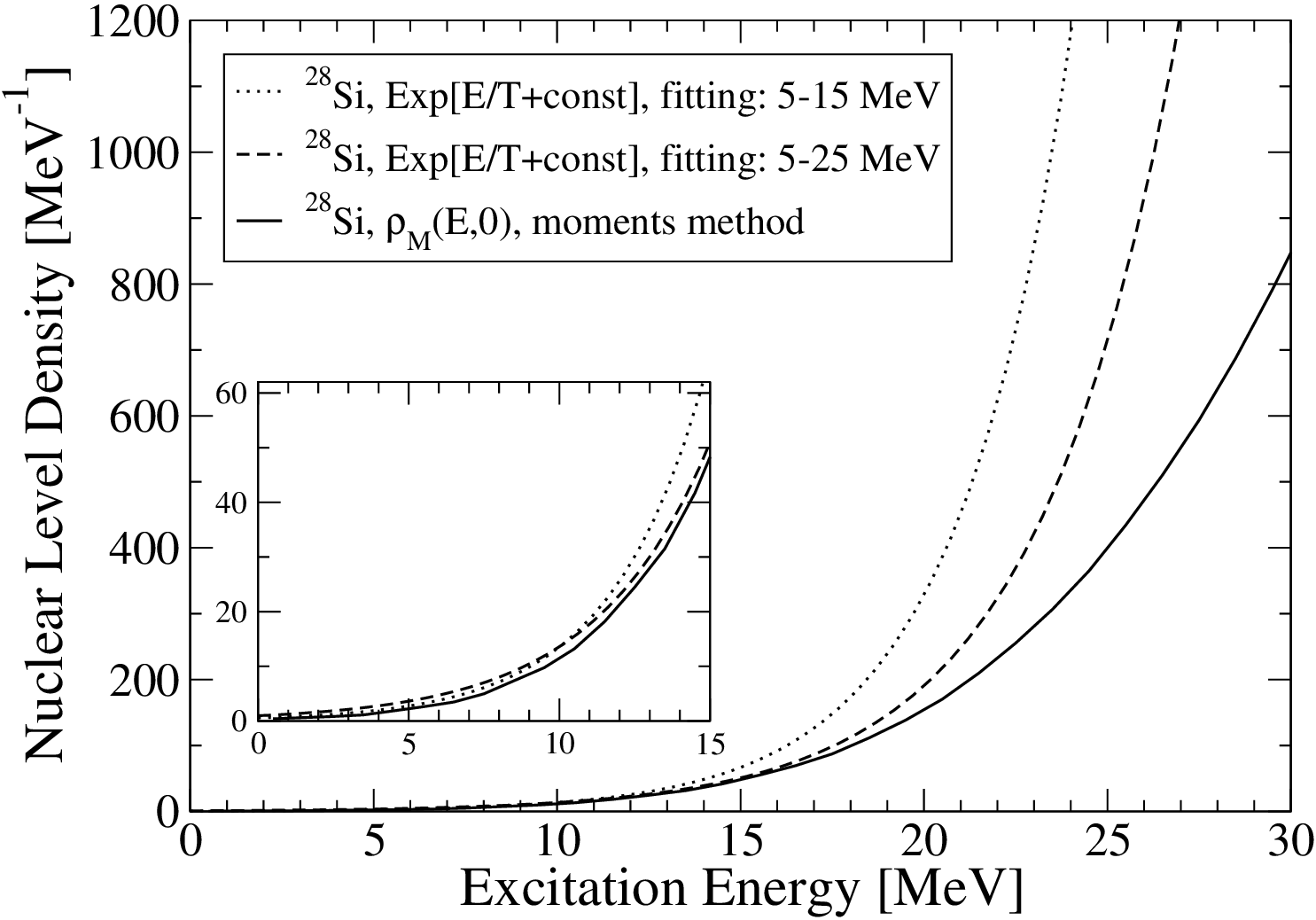}
\caption{Comparison between  global level densities for $^{28}$Si
calculated with the moments method (the solid curve) and using the constant temperature for 5-15 MeV (the dotted curve) and for 5-25 MeV (the dashed curve) energy range; fits according to Eq. (\ref{19}). The inset presents the enhanced low-energy region.}\label{fig17}
\end{minipage}\qquad
\begin{minipage}[t]{.48\textwidth}
\includegraphics[width=1.00\linewidth]{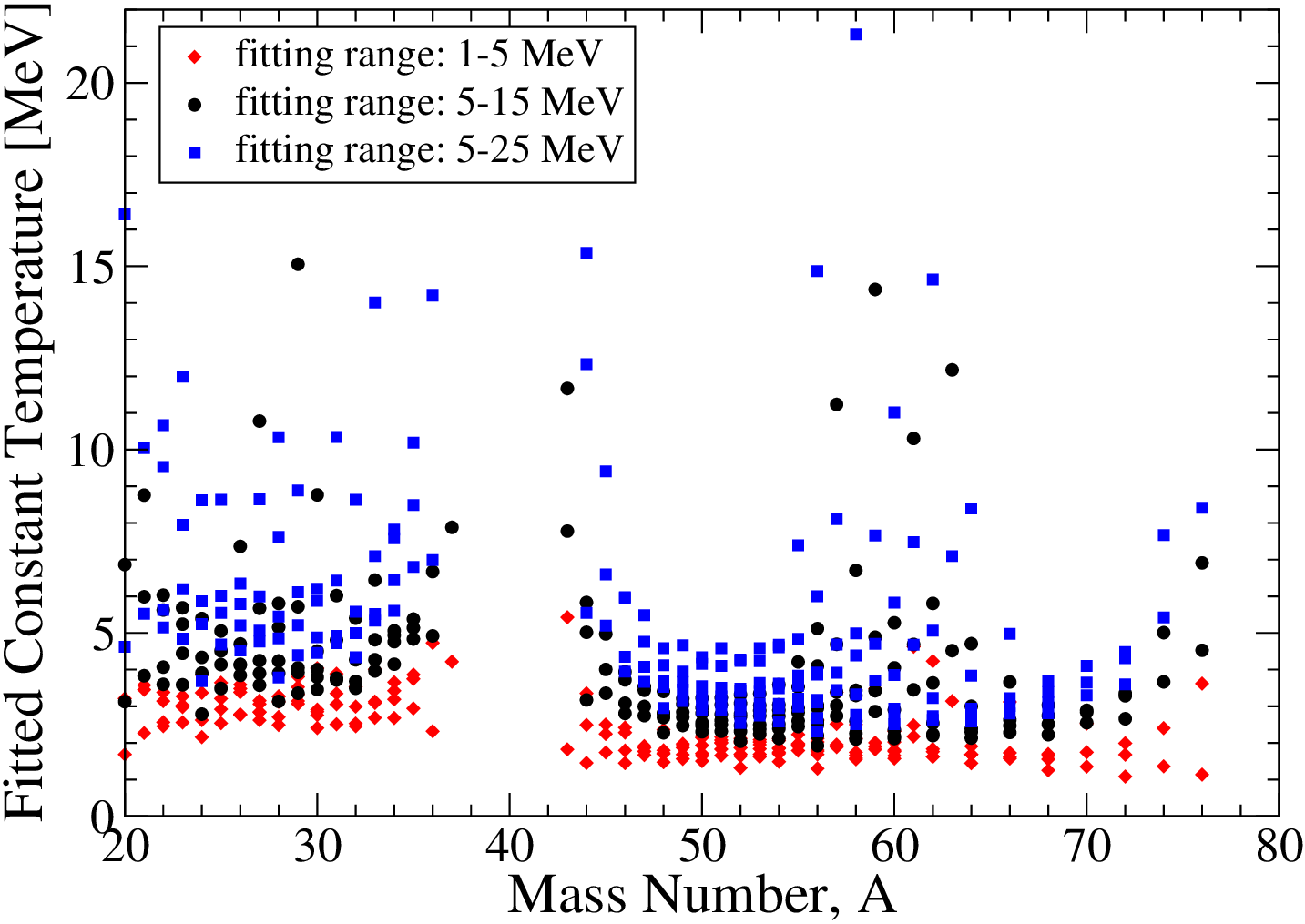}
\caption{(Color online.) Constant temperature fitted for $sd$ and $pf$ shell nuclei. The diamonds (red color) present the  fitting range 1-5 MeV, the circles (black color) correspond to 5-15 MeV and the squares (blue color) correspond to 5-25 MeV fitting ranges.}\label{fig18}
\end{minipage}
\end{figure*}

\subsection{Back-shifted Fermi-gas model}

In this paper we do not discuss in many details the nuclear pairing correlations and its importance for the nuclear level densities leaving this for the future consideration. One of the standard phenomenological way to account for pairing correlation is to use the
back-shifted Fermi-gas formula (BSFG) \cite{gilbert65,brancazio69},
\begin{equation} \label{16}
\rho_{BSFG}(E) =\,\frac{\sqrt{\pi}\,\bar{a}}{12\,[a(E-\delta)]^{5/4}}\,e^{2\sqrt{a(E-\delta)}},\\
\end{equation}
where the excitation energy is shifted by the pairing energy parameter $\delta$. This introduces  a new free parameter that should be fitted alongside with the level density parameter $a$. The pairing energy parameter $\delta$ is in general different for even-even, odd-odd, and even-odd nuclei due to the formation of Cooper pairs, and does not necessarily coincide with the corresponding pairing gap parameters $\Delta$ or $2\Delta$.

Fig. \ref{fig14} shows the sensitivity of the level density parameter $a$ to inclusion of the energy shift $\delta$. The quantity plotted along $y$-axis is the relative change in $a$ if we shift the excitation energy by $\delta$,
\begin{equation}\label{17}
100\% \times \frac{a-a_{BSFG}}{a},
\end{equation}
where $a_{BSFG}$ is the level density parameter fitted using Eq.(\ref{16}) with a fixed value of the pairing energy $\delta$. We can see that the $a$ parameter varies only in a $\pm 6\%$ range when the shift changes from $\delta=-1$ MeV (presented by the empty circles) to  $\delta=1$ MeV (filled circles).

\subsection{Constant temperature model}

The model of the energy dependence of the level density different from the Fermi-gas phenomenology (\ref{12}) being suggested long ago \cite{gilbert65,brancazio69} gradually  becomes  popular among practitioners. It is assumed that the level density, at least up to 10 MeV excitation energy, and maybe even higher \cite{voinov09, guttormsen13,guttormsen14}, can be described by the constant temperature $T$. This temperature  is the single parameter defined, in the simplest version, according to the thermodynamics as
\begin{equation}
T= \,\left[\,\frac{d\ln\rho(E)}{dE}\,\right]^{-1}.                                     \label{18}
\end{equation}
The philosophy behind this approach is usually explained \cite{moretto14} in terms of the first-order phase transition that goes through the latent heat at fixed temperature. Although typically
this assumes the melting of the Cooper pairs but in fact one can also talk about other types of correlated structures which are undergoing something similar to the liquid-gas phase transition
or even the first stage on the road to multifragmentation. In a more detailed description, the effective temperature parameter can be different for the classes of states with different quantum numbers, although such a generalization does not look well from the viewpoint of the
thermal equilibrium between various degrees of freedom. Such an effective temperature parameter (plus a corresponding constant) could be fitted in a reasonable energy range to represent the partial (with certain spin and parity) or total nuclear level densities as
\begin{equation} \label{19}
\begin{aligned}
\ln [ \rho(E,J) ] = \frac{E}{T_J} + \mbox{constant},\\
\ln [ \rho(E) ] = \frac{E}{T} + \mbox{constant}.
\end{aligned}
\end{equation}

\begin{figure*}
\begin{tabular}{cc}
\includegraphics[width=0.44\textwidth]{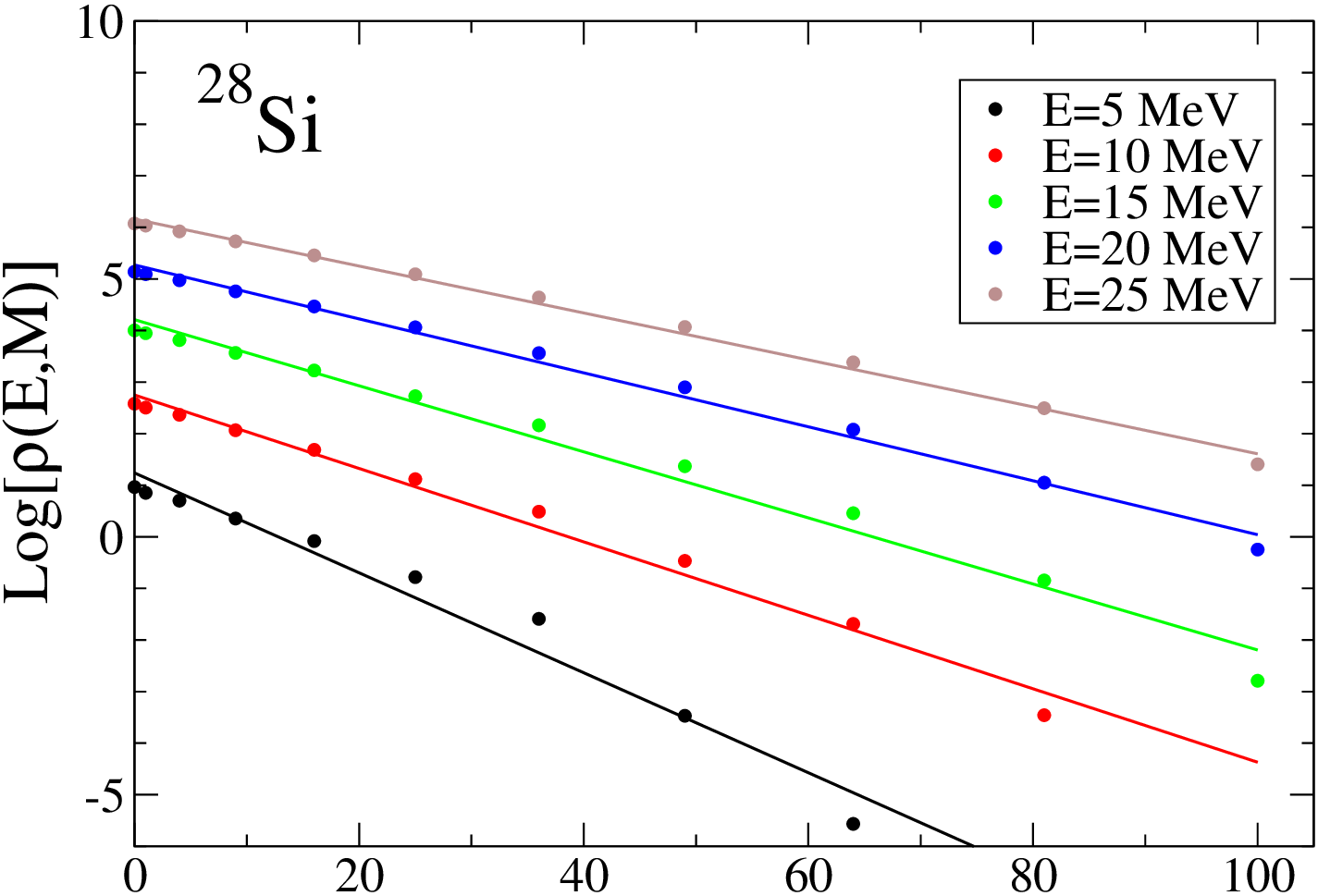} \hspace*{0.1cm} &
\includegraphics[width=0.44\textwidth]{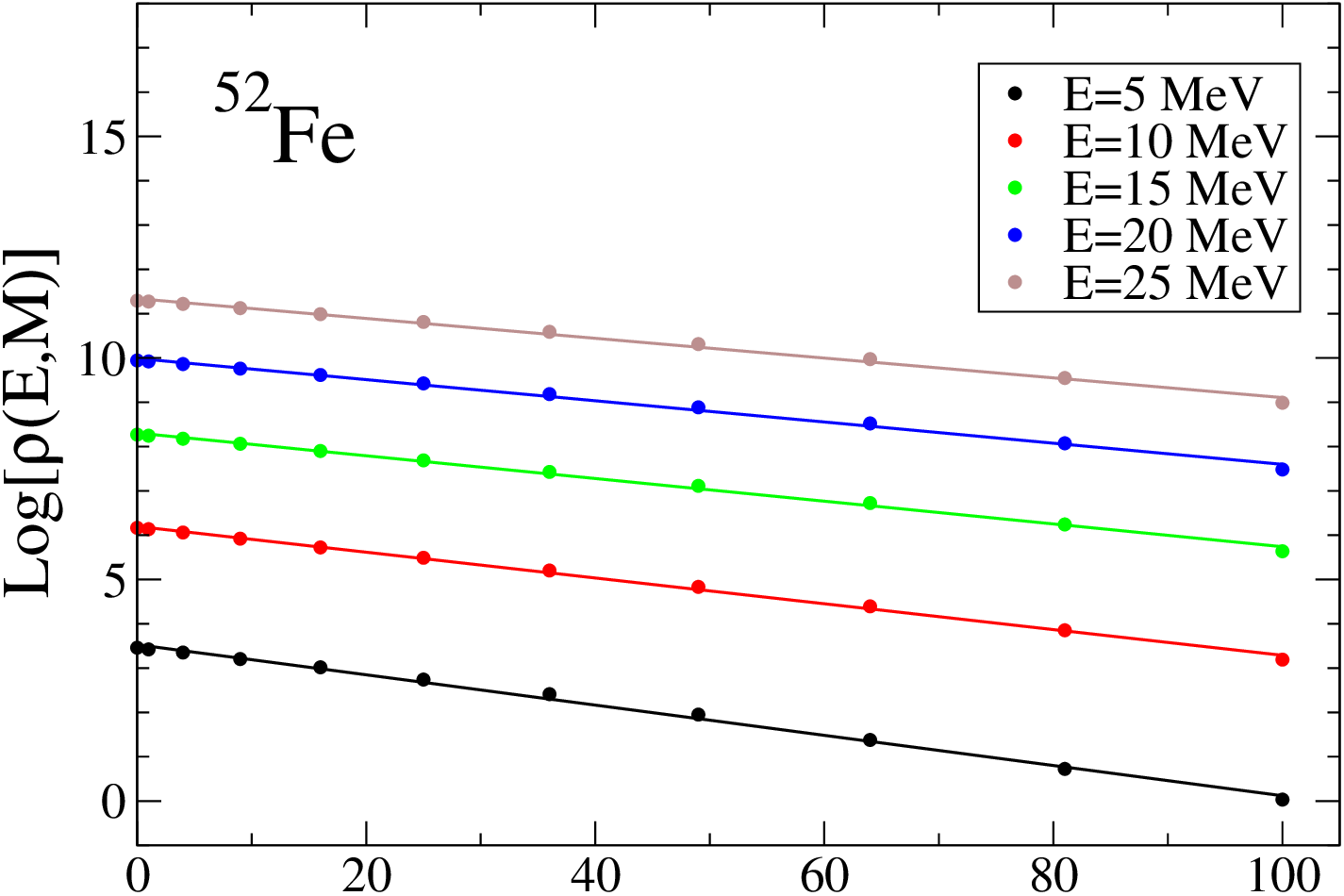} \\
\vspace*{0.5cm}\\
\includegraphics[width=0.44\textwidth]{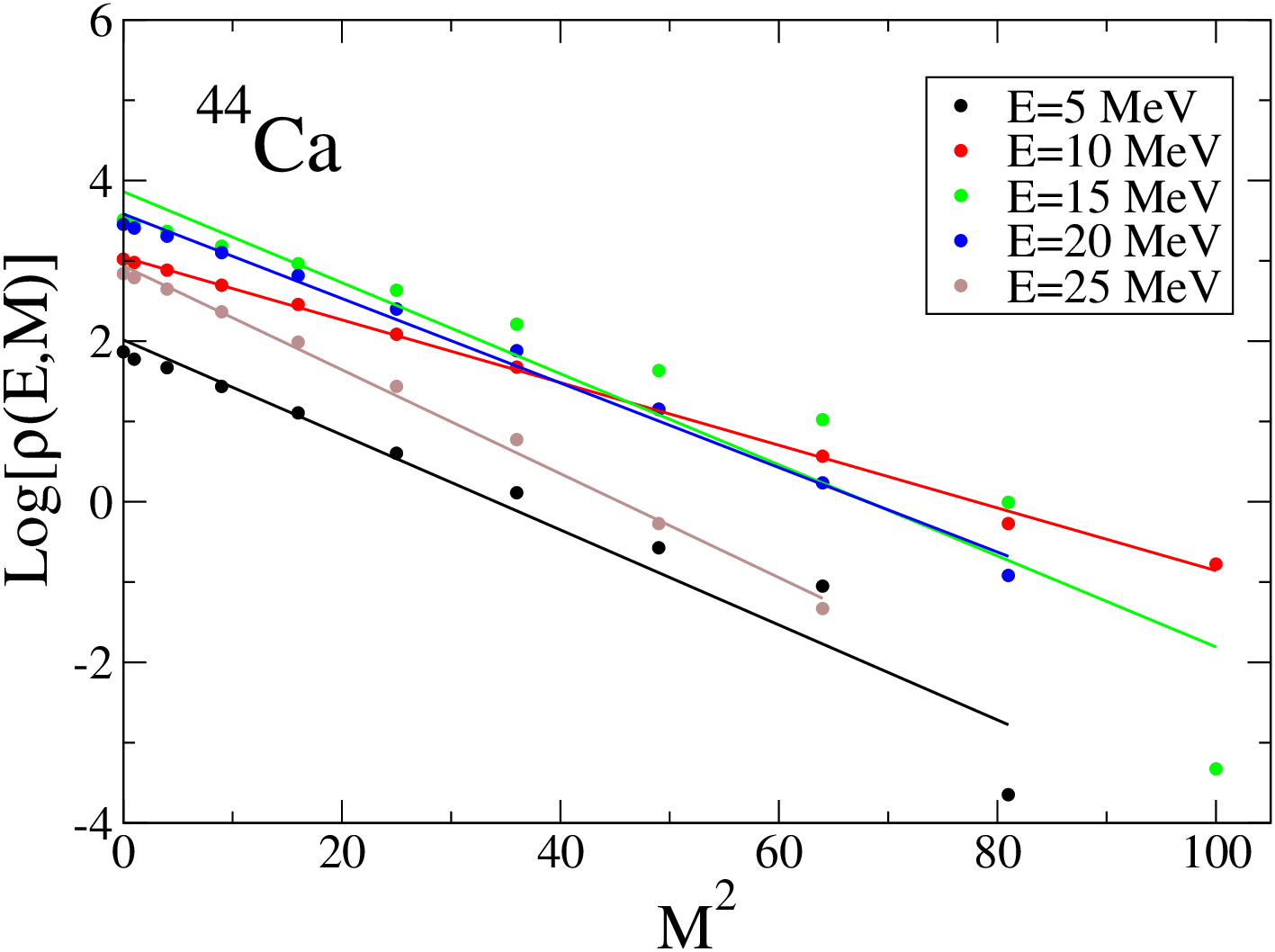} \hspace*{0.1cm} &
\includegraphics[width=0.44\textwidth]{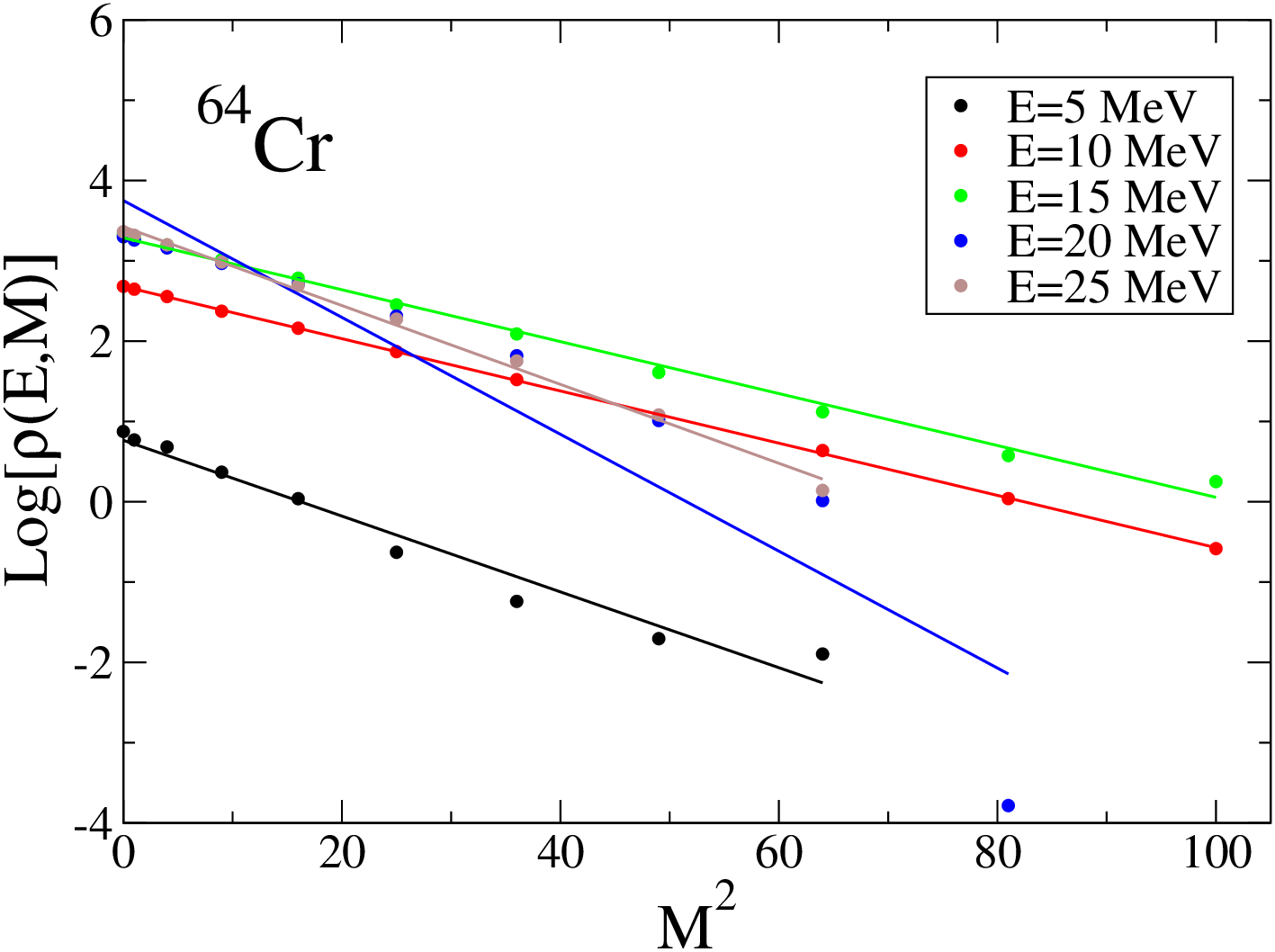} \\
\end{tabular}
\caption{(Color online.) Logarithm of level density, $\ln[\rho(E,M)]$, versus $M^2$, Eq. (\ref{23}), calculated for $^{28}$Si, $^{52}$Fe, $^{44}$Ca, and $^{64}$Cr.  Different colors present different excitation energies: 5 MeV (black), 10 MeV (red), 15 MeV (green), 20 MeV (blue), and 25 MeV (brown). Dots correspond to the moments method calculations, the solid lines are  linear interpolations, see the text.\\}
\label{fig19}
\end{figure*}

It is immediately clear that globally the constant temperature model cannot be compatible with our shell-model calculations. In the truncated orbital space the global level density will always look as a Gaussian with the effective temperature (\ref{18})
\begin{equation}
T_{{\rm eff}}=\,\frac{\sigma_{E}^{2}}{E_{c}-E},                                        \label{20}
\end{equation}
changing with energy from positive to negative values on different sides of the centroid energy
$E_{c}$. Here $\sigma_{E}$ is the effective width of the Gaussian that reflects the summed contribution of all components of interactions. However, at relatively low energy the level density
can grow approximately exponentially effectively resulting in an approximately constant temperature.

Here we give a couple of examples showing the exponential fit to the level density of different $j$-classes and global level density. Fig. \ref{fig15} shows how the actual temperature (see Eq. (\ref{18})) calculated by the moments method for $^{28}$Si and $^{56}$Fe depends on the excitation energy. The temperatures calculated for certain spins $J$ and for all spins (the total density) are not constants, they increase with the excitation energy as suggested in Eq. (\ref{20}). The corresponding constant temperature fit of Eq. (\ref{19}) performed for the $sd$ and $pf$ nuclei is presented in Fig. \ref{fig16}. The effective constant temperatures in the figure depend on the range of the excitation energies where the fit was performed, $-$ the greater the excitation energy the higher the effective constant temperature.

\begin{figure*}
\begin{tabular}{cc}
\includegraphics[width=0.45\textwidth]{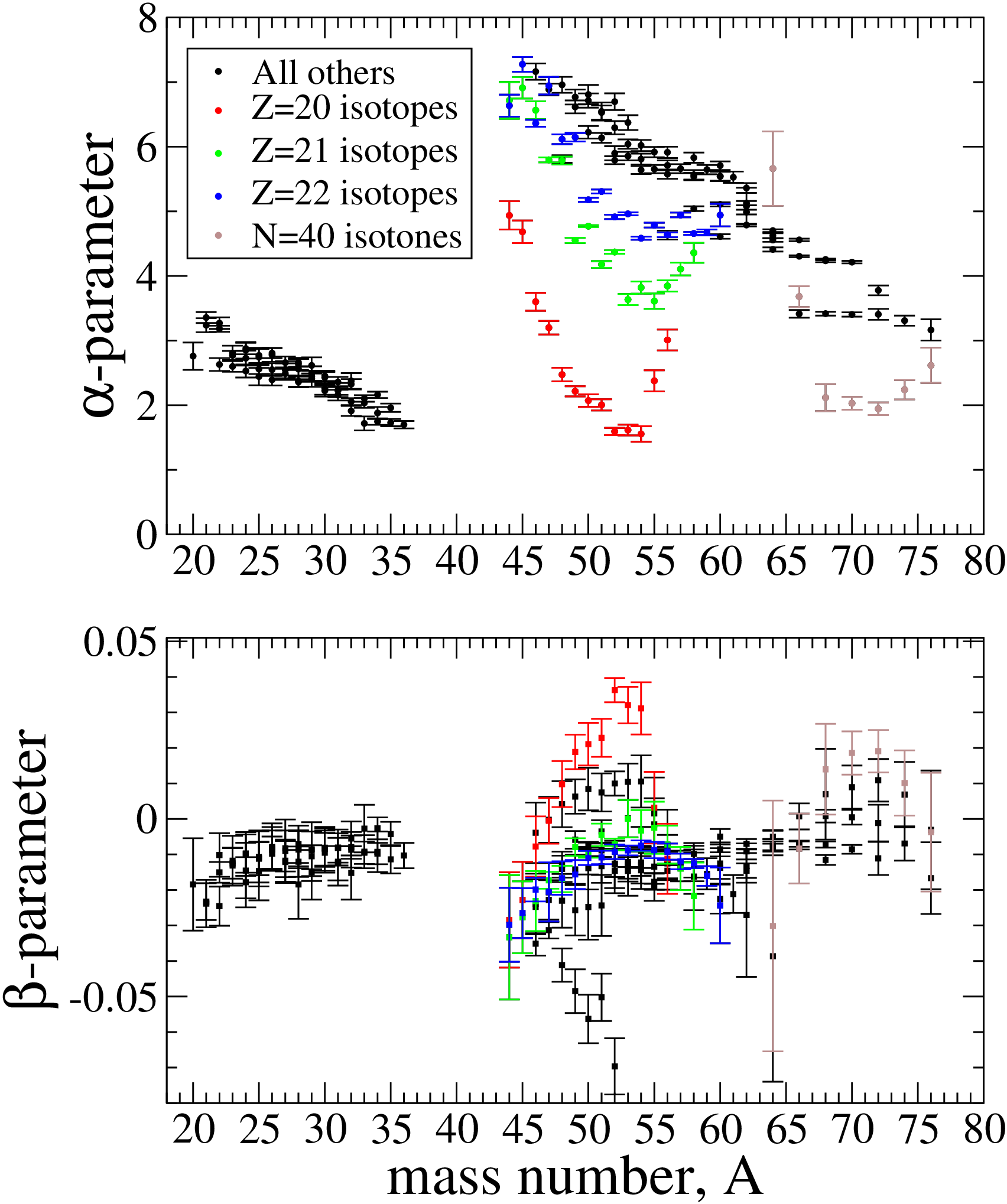} \hspace{0.1pc} &
\includegraphics[width=0.45\textwidth]{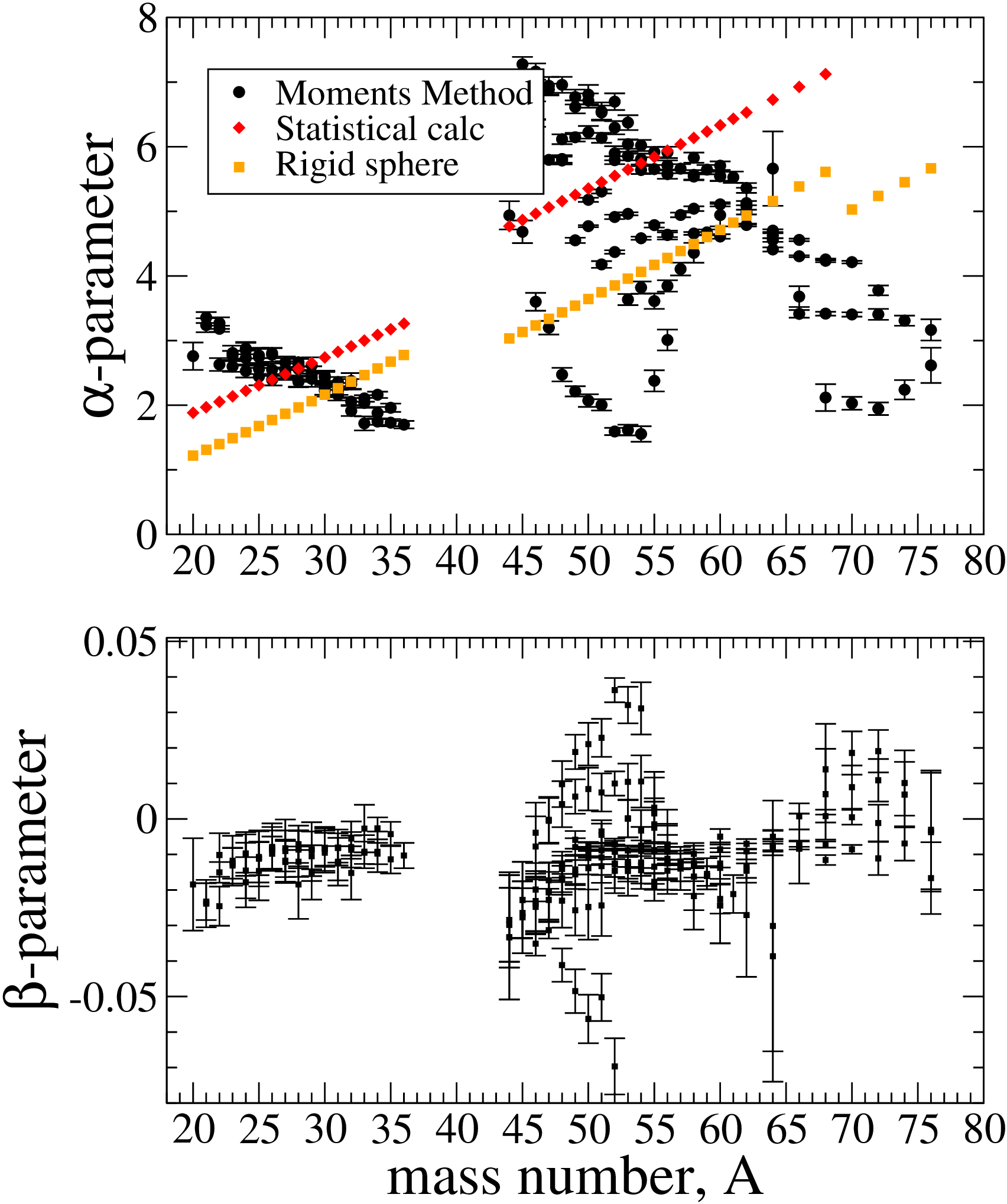} \\
\end{tabular}
\caption{(Color online.) Interpolation of the spin cut-off parameter $\sigma^2$,  Eq. (\ref{24}). Parameters $\alpha$ and $\beta$ are shown for different $sd$ and $pf$ nuclei. Left panel: different colors present different isotopes or isotones. Right panel: moments-method calculation with interpolation (black circles), statistical calculation (orange diamonds), rigid sphere approximation (yellow squares).\\}
\label{fig20}
\end{figure*}

We note that the majority of the fitted temperatures in Figs. \ref{fig15} and \ref{fig16} are concentrated in small regions near 2-5 MeV, while there are some exceptional cases of unreasonably high temperatures of 10-20 MeV that correspond to the nuclei with complete or almost complete shells for one or two sorts of nucleons when the Fermi gas approximation is obviously invalid. Finally, Fig. \ref{fig17} shows how good the constant temperature approximation is. The dotted curve (fitting energy range is 5-15 MeV) and the dashed curve (the range is 5-25 MeV) present the corresponding constant-temperature level densities for $^{28}$Si. We can see that these densities work pretty well inside the energy interval where they were fitted compared to the ``exact" density calculated within the moments method. As the excitation energy increases the constant-temperature densities stop working going too high very fast. Fig. \ref{fig18} shows the fitted temperatures when different energy ranges are used: red diamonds present 1-5 MeV fitting range, black circles and blue squares present 5-15 MeV and 5-25 MeV energy ranges correspondingly. It is natural that at higher excitation energies the effective temperature increases thus reducing the rate at which the density grows, see Eqs. (\ref{19}) and (\ref{20}).

\section{Spin cut-off parameter}

The distribution of levels with certain values of global nuclear constants of motion (angular momentum, parity, isospin) is of special interests in all applications. Our method directly supplies the required information for every set of those exact quantum numbers. The standard phenomenological approach to the angular momentum dependence of the level density assumes the random angular momentum coupling as a diffusion process in the space of projections $M$. The total projection results from the random walk, and the fraction of states with a given projection $M$ is Gaussian,
\begin{equation}
\frac{\rho(E,M)}{\rho(E)}=\,\frac{1}{\sqrt{2\pi\sigma^{2}}}\,
e^{-M^{2}/2\sigma^{2}}.                                                            \label{21}
\end{equation}
Then, as it was mentioned earlier, the density of states with a given value of $J$ is just a difference
\begin{equation}
\rho_{J}(E)=\rho(E,M=J)-\rho(E,M=J+1).                                                 \label{22}
\end{equation}

Assuming this random angular momentum coupling we expect the linear $M^{2}$-dependence of the logarithm of the level density,
\begin{equation}
\ln[\rho(E,M)]=\ln[\rho(E,M=0)]-\,\frac{M^{2}}{2\sigma^{2}}.              \label{23}
\end{equation}
The top two graphs of Fig. \ref{fig19} show the $M^{2}$-dependences of  $\ln[\rho(E,M)]$ at different energies for $^{28}$Si and $^{52}$Fe. The lines correspond to the best linear interpolation of this logarithm. In these examples we see a very good linear behavior and the smooth dependence on excitation energy. This can be interpreted as an evidence for random coupling of angular momenta of individual particles.

The situation is different in two next examples, the bottom graphs in Fig. \ref{fig19}, $^{44}$Ca and $^{64}$Cr. Here we do not have a regular energy dependence and, therefore a clearly defined parameter $\sigma$. This is what could be expected from physical arguments. The whole idea of the Gaussian random walk in the angular momentum space breaks down here because of the isospin limitations. The nucleus $^{44}$Ca in the shell-model description has only four identical $f_{7/2}$ neutrons which allow for the isospin $T=2$ only and for the interaction in the particle-particle channel with the total isospin $T=1$.  Therefore many values of the total spin are forbidden. The second nucleus, $_{24}^{64}$Cr$_{40}$, in the $pf$-shell model has only four
valence protons with the same limitations of the angular momentum coupling. In both cases, it is hard to expect the requirements of the random spin coupling to be satisfied.

The value of the spin cut-off parameter $\sigma$ can be extracted from the curves as in Fig. \ref{fig19} where we observe a good linear behavior. Using the thermodynamic language, we expect this parameter to be proportional to temperature, or, for Fermi-gas, to the square root of energy. The corresponding  parameterization can be taken as
\begin{equation}
\sigma^{2}=\alpha\sqrt{E}\,(1+\beta E).                                            \label{24}
\end{equation}
The coefficient $\alpha$ can be taken \cite{ericson60} from Fermi gas statistical mechanics as $\propto \nu_F T\langle M^{2}\rangle$, where $\nu_F$ is the single-particle level density at the Fermi surface, or assuming the angular momentum corresponding in average to the rigid-body rotation with the moment of inertia $\propto TA^{5/3}$. The results of the shell-model calculations are shown in Fig. \ref{fig20}. We define the parameters $\alpha$ and $\beta$, Eq. (\ref{24}), from the energy region 5-25 MeV. Comparing two groups of nuclei, $sd$-shell and $pf$-shell, we indeed see the average growth of the spin-cut off parameter for two representative groups as proportional to $A^{5/3}$. It is impossible here to make a selection between the statistical estimate of the spin cut-off parameter and the estimate from the moment of inertia as both of them, being too crude to reflect shell efects inside each group which are certainly present, do not agree with the $A^{5/3}$ estimate and require a more detailed analysis. The constant $\beta$ from Eq. (\ref{24}) is small but, at least for the $pf$-nuclei, may also contain some shell effects.

\section{Conclusion}

In this article we collected, explained and overviewed the first results of the improved method for statistical calculation of the nuclear level density for a given shell-model Hamiltonian. The method is physically based on the chaotization of the intrinsic dynamics by the inter-particle interactions.
In practice, one needs to calculate only the lowest moments of the Hamiltonian partitioned in terms of mean-field configurations. The first two moments turn out to be sufficient for the full agreement of the found level density with the result of the exact diagonalization as it was checked by the cases when such full diagonalization was technically possible. The serious improvements compared to the  previous attempts in the same direction include the use of the
finite-range Gaussian distributions and of the recurrence relation for eliminating the spurious states. We did not discuss the determination of the ground state energy that is necessary for the appropriate positioning of the level density. There are special methods for doing this, including the exponential extrapolation also based on the chaotic properties of remote highly excited states \cite{exp}. The shell-model level density can be calculated in any specific class of global constants of motion (proton and neutron numbers, total spin, parity and isospin) as a function of excitation energy. This is essentially what is needed for practical applications to nuclear reactions including those in astrophysics.

The main conclusion that can be drawn from this experience is that the shell-model level density that results from the statistical calculation is a smooth function of excitation energy in all classes of quantum numbers containing a considerable number of states allowed by the truncation of the orbital space (of course, the state with the maximum possible total spin is frequently just unique). Giant oscillations of the level density predicted by the calculations based on the mean-field combinatorics are almost completely erased by the presence of incoherent collision-like interactions which  usually remain outside of the mean-field models or parametrizations with the so-called collective enhancement. Taking into account all components of residual interactions, coherent (such as pairing) and incoherent, is necessary for the adequate description.

The comparison with phenomenological Fermi-gas approaches, including the models with constant temperature, shows that, being less theoretically justified than the full direct calculation, in many cases they are nevertheless quite reasonable for practical use. The calculation of the spin dependence of the level density and of the relevant spin cut-off parameter is more sensitive to
assumptions, and there are cases when it is not in good agreement with exact results. The model of constant temperature, in our opinion, can be applied at relatively low excitation energy but, most probably, it reflects the general process of chaotization the dynamics rather than just breaking of Cooper pairs. Certainly, the accumulation of experimental data and new applications of the statistical method are necessary for better understanding the underlying physics.

The whole approach unavoidably suffers from the general problems of the shell model. It is possible to believe that the results will not be sensitive to the specific versions of the shell-model Hamiltonian as soon as this choice agrees well with the low-lying spectroscopy. However, the space truncation provides a natural limitation for the applications of all such methods. The space can be expanded (and many-body residual interactions can be included) paying the price of longer computational time that can be cut off by the parallelization. But anyway for any choice of finite space there is a natural limit of applicability. Luckily enough, it seems that this limitation is not essential for many astrophysical applications. More theoretical work is necessary for understanding the resonance density for the states deeply in the continuum which again might not be critical for a typical stellar temperature when the resonance states under consideration still are quite narrow.

The whole development of the method was done in collaboration with M. Horoi. The discussions with B.A. Brown are acknowledged. The work on level density was supported by the NSF grants PHY-1068217 and PHY-1404442.

\end{document}